\documentclass[runningheads]{sty/llncs}
\usepackage[T1]{fontenc}
\usepackage[usenames,dvipsnames]{xcolor}
\usepackage{float}
\usepackage{graphicx}
\usepackage{amsmath,amstext,amssymb}
\usepackage{booktabs}
\usepackage{tabularx}
\usepackage{multirow}
\usepackage{siunitx}
\usepackage{listings}
\makeatletter
\AtBeginDocument{%
  \let\c@lstlisting\c@figure
}
\makeatother
\usepackage{ragged2e}
\usepackage{wrapfig}
\usepackage{textcomp}
\usepackage{makecell}
\usepackage{subcaption} 
\usepackage{rotating}
\usepackage{algorithm2e}
\usepackage{fancyvrb}
\usepackage[most]{tcolorbox}
\usepackage{fontawesome5}
\usepackage{tikz}
\usetikzlibrary{positioning,arrows.meta,shapes.multipart,calc,fit,backgrounds}
\usepackage{xspace}
\usepackage{xstring}
\usepackage{enumitem}
\usepackage{balance}
\usepackage{microtype}
\usepackage{hyperref}

\AtBeginDocument{%
  \hypersetup{colorlinks=true, citecolor=blue, linkcolor=blue, urlcolor=blue}%
  \urlstyle{sf}%
}

\definecolor{bblue}{HTML}{4F81BD}
\definecolor{rred}{HTML}{C0504D}
\definecolor{ggreen}{HTML}{9BBB59}
\definecolor{ppurple}{HTML}{9F4C7C}
\definecolor{darkgray}{rgb}{0.66,0.66,0.66}
\definecolor{gray}{RGB}{136,136,136}
\definecolor{dkgreen}{rgb}{0,0.6,0}
\definecolor{mauve}{rgb}{0.58,0,0.82}
\definecolor{comment-red}{rgb}{0.8,0,0}

\newlength\tbspace
\newcolumntype{C}{c<{\hspace{\tbspace}}}

\SetCommentSty{mycommfont}

\makeatletter
\newcommand\footnoteref[1]{\protected@xdef\@thefnmark{\ref{#1}}\@footnotemark}
\makeatother
\newcommand{\DefMacro}[2]{\expandafter\newcommand\csname rmk-#1\endcsname{#2}}
\newcommand{\UseMacro}[1]{\csname rmk-#1\endcsname}

\newcommand{\Space}[1]{}

\makeatletter
\newcommand{\labitem}[2]{%
  \def\@itemlabel{\textbf{#1}}%
  \item
  \def\@currentlabel{#1}\label{#2}}
\makeatother

\newcommand{\RN}[1]{\textup{\uppercase\expandafter{\romannumeral#1}}}
\SetKwProg{Fn}{Function}{}{}

\newcommand{\PP}[1]{%
  \vspace{2px}%
  \noindent{\bf \IfEndWith{#1}{.}{#1}{#1.}}%
}

\newcommand{\heading}[1]{\vspace{2pt}\noindent{\underline{\textsc{#1}}}}

\begin{document}

\title{Effective and Efficient Threat Hunting with Small Language Models}
\titlerunning{Effective and Efficient Threat Hunting with Small Language Models}
\author{Saleha Muzammil\inst{1}\fnmsep\thanks{These authors contributed equally to this work.}\textsuperscript{(\faEnvelope)} \and
Rahul Reddy\inst{1}\fnmsep\textsuperscript{\ensuremath{\star}} \and
Vishal Kamalakrishnan\inst{1} \and
Hadi Ahmadi\inst{2} \and
Wajih Ul Hassan\inst{1}}
\authorrunning{Muzammil et al.}
\institute{University of Virginia, Charlottesville, VA, USA\\
\email{evz4sc@virginia.edu, dqb5ty@virginia.edu,}\\
\email{vishkamalk@gmail.com, hassan@virginia.edu} \and
Corvic, USA\\
\email{hadi@corvic.ai}}
\maketitle

\begin{abstract}
Analysts in Security Operations Centers query massive telemetry streams using Kusto Query Language (KQL), but writing correct KQL demands specialized expertise that bottlenecks scaling security teams. We investigate how Small Language Models (SLMs) can enable accurate, cost-effective translation from natural language queries (NLQs) to KQL. We propose a three-knob framework spanning prompting, fine-tuning, and architecture. First, we adapt NL2KQL for SLMs with lightweight retrieval and introduce error-aware prompting that targets common parser failures with a handful of mined tips, at a fraction of the tokens KQL's full rule set would require. Second, we apply LoRA fine-tuning with rationale distillation, augmenting each NLQ--KQL pair with a brief chain-of-thought to transfer teacher reasoning. This yields an informative negative result, as neither variant surpasses targeted prompting. Third, we propose a two-stage architecture pairing an SLM drafter with a low-cost LLM judge for schema-aware refinement. We evaluate nine models (five SLMs, four LLMs) on syntax correctness, semantic accuracy, table selection, filter precision, latency, and token cost. On Microsoft's NL2KQL Defender Evaluation dataset, our two-stage approach reaches 0.987 syntax and 0.906 schema-valid (``semantic'') accuracy, exceeding every baseline we run under equivalent infrastructure, and it generalizes to independently authored queries over the same schema (0.964 syntax, 0.831 schema-valid). The only baselines within 0.05 schema-valid are NL2KQL+GPT-4o (0.878) and NL2KQL+GPT-5 (0.861), which cost \$2.998 and \$2.018 for 230 queries against \$0.213 for ours, a 9.5--14$\times$ reduction at matched accuracy. These results establish SLMs as a practical foundation for natural-language querying in security operations.

\keywords{Security Operations \and Small Language Models \and Natural Language to Query Generation}
\end{abstract}

\section{Introduction}
\label{s:intro}

In Security Operations Centers (SOCs), analysts face overwhelming volumes of event logs, network traffic, and threat intelligence feeds \cite{Bargan_2024}, and report false-positive rates high enough that nearly every alarm must be validated by hand \cite{Alahmadi}. A substantial body of work attacks this burden upstream of the analyst, either by triaging alerts automatically before they are ever read \cite{nodoze2019,rapsheet2020} or by detecting intrusions directly from system audit logs \cite{inam2022sok,flash2024}. The investigation that follows an alert, however, still falls to the analyst, and hunting through telemetry at this scale is itself a systems challenge \cite{swift2020}. To investigate these massive logs, they rely on query languages such as Sigma, Elastic EQL, and KQL \cite{SigmaHQ,Elastic_Query_Language_Documentation,Kusto-Query-Language}, but translating natural-language intent into correct queries over large, evolving schemas remains a major bottleneck.

KQL \cite{Kusto-Query-Language} is among the most widely used of these in enterprise settings. Integrated into Microsoft Sentinel and Defender, it is a domain-specific, read-only language that lets analysts parse millions of log rows efficiently and extract relevant events. However, its expressiveness and non-trivial semantics make authoring accurate queries challenging for non-experts, especially during time-sensitive investigations.

Automated KQL generation can accelerate investigations by producing precise queries aligned with an analyst's intent, and LLMs have been applied across cybersecurity tasks such as vulnerability detection and penetration testing \cite{PentestGPT,VulDetect}. A natural language query (NLQ) is an analyst's request stated in plain English, such as ``which devices connected to 89.12.55.1 in the last week''; Figure~\ref{fig:kql-example} shows the KQL that answers it. However, LLM-centric pipelines for translating an NLQ into KQL can be costly, latency-sensitive, and difficult to deploy under enterprise data-governance constraints. We refer to this task as NLQ-to-KQL translation throughout.

As an alternative, we turn to SLMs, defined following \cite{Belcak_Heinrich_Diao_Fu_Dong_Muralidharan_Lin_Molchanov} as language models that fit on a common consumer device and serve one user's requests at practical inference latency; that work treats most models below 10 billion parameters as qualifying, which is the bound we adopt. SLMs offer lower latency, memory, and operational costs while still achieving accuracy comparable to LLMs in some domain-specific tasks \cite{Belcak_Heinrich_Diao_Fu_Dong_Muralidharan_Lin_Molchanov}. Despite this promise, there has been no systematic study of SLMs for NLQ-to-KQL translation in realistic, schema-rich settings.

\begin{figure}[!t]
\centering
\begin{minipage}{0.95\linewidth}
\begin{lstlisting}[language=SQL,
    basicstyle=\ttfamily\scriptsize,
    keywordstyle=\color{blue}\bfseries,
    commentstyle=\color{gray}\itshape,
    stringstyle=\color{teal},
    showstringspaces=false,
    frame=single,
    caption={Example KQL query that retrieves device IDs which last connected from IP address 89.12.55.1 within the last 7 days. The query references multiple columns in the \texttt{DeviceNetworkEvents} table schema, including \texttt{Timestamp}.},
    captionpos=b,
    label={fig:kql-example}]
DeviceNetworkEvents
| where Timestamp >= ago(7d)
| where ActionType == 'ConnectionSuccess'
| summarize arg_max(Timestamp, LocalIP) by DeviceId
| where LocalIP == "89.12.55.1"
| project DeviceId
\end{lstlisting}
\end{minipage}
\vspace{-3ex}
\end{figure}

This work fills that gap through three orthogonal enhancement knobs, namely prompting, fine-tuning, and two-stage architecture design, together with a comprehensive empirical study spanning accuracy, latency, and cost. We adapt the NL2KQL architecture for SLMs by replacing heavy components with lightweight alternatives while preserving the prompting, retrieval, and refinement structure so that results are both comparable and cost efficient. We further mine the KQL parser's error distribution to derive prompting tips that target the failures SLMs actually make, rather than supplying the language's full rule set.

Beyond prompting and fine-tuning, we introduce a two-stage architecture for KQL generation. In the first stage, an SLM produces an initial KQL draft efficiently. In the second stage, a lightweight LLM performs constrained refinement by applying minimal corrections to the SLM draft using schema information and parsing feedback. This design assigns each model the task it does best: the SLM handles bulk generation cheaply, while the LLM applies schema-aware judgment only at the final step. Efficiency throughout this paper means cost and parameter efficiency at a fixed accuracy bar. The two sequential stages make per-query latency predictable but not minimal, and we report where that trade binds in \S\ref{s:eval}.

Our experiments span baseline performance, prompting strategies, fine-tuning, and architectural design. We evaluate nine models, five SLMs and four LLMs (listed in \S\ref{s:eval}), on 230 NLQ-KQL pairs from Microsoft's NL2KQL Defender Evaluation Dataset~\cite{NL2KQL_Eval}, and test generalization on 83 independently authored queries from Sentinel-Queries~\cite{reprise99}.

LLMs generally achieve high syntactic accuracy but variable schema validity (our ``semantic score'', defined in \S\ref{s:eval}), while SLMs perform poorly zero-shot, with schema validity near 0 and table and filter-column scores below 0.1. Targeted prompting closes much of that gap. LoRA fine-tuning, with or without rationale distillation, does not, a negative result we take up in \S\ref{s:eval}. Our two-stage SLM-Oracle architecture is the strongest configuration we report and the cheapest at its accuracy level, and it carries that structural competence over to independently authored queries, though literal-level precision (Filter$_{\text{lit}}$) does not transfer as reliably (\S\ref{sec:general}).

Our paper makes the following main contributions:
\begin{itemize}[leftmargin=*]
\item We present the first systematic evaluation of open-weight SLMs (under 10B parameters) for NLQ-to-KQL translation in SOC workflows.
\item We show that schema-aware prompting substantially improves the correctness of SLM-generated KQL queries, providing a strong alternative to zero-shot generation.
\item We introduce error-aware prompting, in which a handful of tips mined from the KQL parser's error distribution improve syntactic and semantic correctness at a fraction of the token cost of the full rule set.
\item We propose a two-stage architecture that combines SLM drafting with lightweight LLM refinement. It attains the highest syntax and schema-valid scores of any configuration we report, at 9.5$\times$ and 14$\times$ lower cost, under equivalent infrastructure, than the GPT-5- and GPT-4o-based NL2KQL reimplementations, the only two baselines within 0.05 schema-valid of it.
\end{itemize}
 
\section{Related Work}
\label{s:relwk}




\heading{Natural Language to Query Generation} Converting NLQs to structured queries has been widely studied across domains and languages. Early work \cite{popescu2003towards,li2014nalir} translated NLQs to SQL using rule-based methods and custom heuristics, later mining SQL query logs to cut that manual effort \cite{baik2019bridging}, but extending these techniques to new databases and languages struggles with edge cases \cite{nl2kql}. Neural methods improve portability by casting text-to-SQL as sequence generation with integrated validation. \textit{PICARD} \cite{PICARD} constrains decoding by incrementally parsing each partial output, rejecting tokens that cannot extend to a valid, schema-consistent query. Augmenting the model with generated domain knowledge about the database further improves semantic fidelity \cite{Hong_Yuan_Chen_Zhang_Huang_Huang_2024}. These ideas combine constrained decoding for syntax with schema-aware prompting or retrieval for semantics, directly informing our KQL setting.

\heading{Security Query Languages} These general principles have been extended to domain-specific query languages, especially in security, where KQL is the primary language for threat hunting and log analysis. \textit{NL2KQL} \cite{nl2kql} translates NLQs to KQL by combining embedding-based semantic similarity with few-shot prompting in a modular, end-to-end framework. \textit{Xpert} \cite{X_Pert} takes an incident-centric approach, retrieving similar historical tickets to recommend or compose queries using in-context learning with embedding-based retrieval. However, both systems rely heavily on LLMs, resulting in cost and latency constraints that limit real-time SOC deployment, where analysts need rapid query assistance. In contrast, our approach delegates generation to an SLM and reserves the LLM for a single refinement step.


\heading{Security-Specific LLMs}
LLMs are increasingly applied across SOC tasks and offensive evaluations, but their effectiveness hinges on task-specific scaffolding and verification. \textit{VulDetect} \cite{VulDetect} fine-tunes a pre-trained GPT model to flag vulnerable source-code patterns without the compute overhead of CNN- and LSTM-based detectors. \textit{PentestGPT} \cite{PentestGPT} builds an LLM penetration-testing assistant with modular task support, reporting bounded gains on real targets, highlighting the importance of structured tooling around the model. Although LLMs offer strong code generation capabilities, their cost and latency often make real-time deployment impractical~\cite{Belcak_Heinrich_Diao_Fu_Dong_Muralidharan_Lin_Molchanov}, pushing real-time operations toward SLMs.

\heading{SLM Adoption}
SLMs provide attractive latency and cost profiles for SOC settings but require augmentation to match LLM quality. SuperICL shows that small, locally fine-tuned models can act as plug-ins to larger LLMs, structuring context while the larger model executes, a useful division of labor for SOC workflows \cite{SLM_Ref}. Parameter-efficient fine-tuning further narrows this gap while preserving efficiency, and methods like LoRA and QLoRA enable targeted adaptation on modest hardware \cite{LoRA,Improving_Phishing_Email_Detection,qlora}. Complementary techniques such as schema/value retrieval, few-shot selection, voting over sampled reasoning paths \cite{wang2022self}, and iterative self-feedback \cite{madaan2023self} improve correctness without enlarging model footprint.

\heading{Model Cascades and LLM-as-a-Judge} Our two-stage design belongs to a broader family of systems that route work between models of differing capability to reduce cost at fixed quality. Cost-optimization work for LLM serving shows that predicting output quality before invocation and selecting the cheapest adequate model cuts spend by 40--90\% \cite{Cost_Optimization}, and self-refinement demonstrates that a second pass over a generated draft can correct errors the first pass could not detect \cite{madaan2023self}. The judgment step itself is typically instantiated as an LLM-as-a-Judge, which aligns well with human preferences while exhibiting recognizable biases \cite{zheng2023judging}. Our setting differs in two ways. The cascade is heterogeneous by design rather than routed per input, so every draft is refined and both latency and cost are predictable. The judge is also given the retrieved schema, making its decision a grounded validity check rather than a preference judgment (\S\ref{s:eval}).

\section{Methodology}
\label{s:methodology}

\providecommand{\knobbadge}[2]{{\setlength{\fboxsep}{1.2pt}%
  \colorbox{#1!70!black}{\textcolor{white}{\scriptsize\bfseries\sffamily #2}}}}
\providecommand{\knobhead}[4]{{\textcolor{#1!65!black}{\footnotesize #2}}\; %
  {\footnotesize\bfseries\textcolor{#1!40!black}{#4}}\hfill\knobbadge{#1}{#3}%
  \\[0.7mm]{\color{#1!50!black}\rule{\linewidth}{0.3pt}}\\[-0.4mm]}
\providecommand{\knobfoot}[1]{\par\vspace{0.5mm}{\color{black!60}#1}\par}
\newlength{\knobcardht}\setlength{\knobcardht}{2.95cm}
\providecommand{\knobbody}[1]{\begin{minipage}[t][\knobcardht][t]{\linewidth}#1\end{minipage}}
\SetEnumitemKey{knobitems}{leftmargin=2.8mm, itemsep=0.5mm, topsep=0.8mm,
  parsep=0pt, partopsep=0pt, label=\textcolor{black!40}{\tiny$\bullet$}}

\begin{figure}[!t]
\centering
\begin{tikzpicture}[
  font=\sffamily\scriptsize,
  stage/.style={
    draw=black!45, fill=black!3, rounded corners=1.5pt,
    text width=1.55cm, align=center, inner sep=0.9mm, minimum height=9.5mm
  },
  touched/.style={draw=#1!65!black, fill=#1!8},
  flow/.style={-{Stealth[length=1.6mm,width=1.4mm]}, semithick, draw=black!55},
  hook/.style={dashed, semithick, draw=#1!60!black,
               -{Stealth[length=1.5mm,width=1.3mm]}},
  card/.style={
    draw=#1!60!black, fill=#1!5, rounded corners=2pt,
    text width=3.55cm, align=left, inner sep=1.5mm, anchor=north west,
    minimum height=2.12cm
  }
]

\node[stage] (in) at (0.95,0)
  {\textcolor{black!55}{\faComment}\\NLQ $t$, schema $s$};
\node[stage, touched=bblue] (pr) at (2.95,0)
  {\textcolor{bblue!60!black}{\faAlignLeft}\\prompt assembly};
\node[stage, touched=ggreen] (slm) at (4.95,0)
  {\textcolor{ggreen!55!black}{\faMicrochip}\\base SLM $\mathcal{L}$};
\node[stage] (draft) at (6.95,0)
  {\textcolor{black!55}{\faCode}\\draft $\hat{q}_{0}$};
\node[stage, touched=ppurple, dashed] (orc) at (8.95,0)
  {\textcolor{ppurple!60!black}{\faGavel}\\Oracle refinement};
\node[stage] (out) at (10.95,0)
  {\textcolor{black!55}{\faCheck}\\final KQL $\hat{q}$};

\draw[flow] (in)    -- (pr);
\draw[flow] (pr)    -- (slm);
\draw[flow] (slm)   -- (draft);
\draw[flow] (draft) -- (orc);
\draw[flow] (orc)   -- (out);

\draw[flow, dashed] (draft.north) to[out=50,in=130]
  node[midway, above, font=\sffamily\tiny, text=black!55, inner sep=0.4mm]
  {without $\mathbf{M}$} (out.north);

\node[card=bblue] (cp) at (0.00,-1.62) {
\knobbody{%
  \knobhead{bblue}{\faAlignLeft}{P}{Prompting scheme}
  \begin{itemize}[knobitems]
    \item zero-shot
    \item NL2KQL grounding: top-9 tables, 2 exemplars
    \item error-aware tips
  \end{itemize}
  \vfill
  \knobfoot{prompt varies \;(RQ2--RQ3)}
}
};

\node[card=ggreen] (cf) at (4.05,-1.62) {
\knobbody{%
  \knobhead{ggreen}{\faMicrochip}{F}{LoRA fine-tuning}
  \begin{itemize}[knobitems]
    \item teacher: Gemini 2.0 Flash
    \item 800 CoT$+$KQL pairs (200 held out)
    \item student: DeepSeek Coder 6.7B
  \end{itemize}
  \vfill
  \knobfoot{weights vary \;(RQ4)}
}
};

\node[card=ppurple] (cm) at (8.10,-1.62) {
\knobbody{%
  \knobhead{ppurple}{\faGavel}{M}{Two-stage pipeline}
  \begin{itemize}[knobitems]
    \item SLM drafts, Oracle refines
    \item refine-only vs.\ schema-aware
    \item LLM tokens at last step
  \end{itemize}
  \vfill
  \knobfoot{pipeline varies \;(RQ5)}
}
};

\draw[hook=bblue]   (2.95,-1.62) -- (pr.south);
\draw[hook=ggreen]  (4.95,-1.62) -- (slm.south);
\draw[hook=ppurple] (8.95,-1.62) -- (orc.south);

\end{tikzpicture}
\caption{The three orthogonal enhancement knobs. All configurations share the
same translation pipeline (top): the NLQ $t$ and schema context $s$ are
assembled into a prompt, decoded by a base SLM $\mathcal{L}$, and emitted as
$\hat{q}$. Each knob intervenes at exactly one stage, $\mathbf{P}$ at prompt
assembly, $\mathbf{F}$ at the model weights, and $\mathbf{M}$ by inserting an
Oracle refinement stage after the draft, so the marginal contribution of each
can be measured while the other two are held fixed.}
\label{fig:knob_diagram}
\vspace{-2ex}
\end{figure}

\subsection{Problem Statement}
\label{s:problem}


Three properties make NLQ-to-KQL translation hard. The cost of an incorrect query is asymmetric: a query that filters on the wrong column silently returns no rows, letting a real threat go undetected, while an over-broad query adds to alert fatigue. Correctness itself depends on knowledge most language models lack, since KQL's pipe-based operators differ subtly from SQL and using them correctly requires exact table schemas, column types, and enumerated values. And the failure modes are not self-detectable: a plausible-looking query may reference a non-existent column, apply an operator in prefix rather than infix form, or join tables incorrectly, none of which the model can catch without external validation. These constraints motivate a translation pipeline with explicit schema grounding and correctness validation.

Recall from Section~\ref{s:intro} that an NLQ is a request stated in natural language that the user wishes to have answered in terms of KQL. The set $\mathcal{T}$ represents the set of all possible NLQs. A schema is defined as the
columns that are associated with each single table in KQL. The set $\mathcal{S}$ represents the set of all possible schema contexts that are associated with tables in KQL. Lastly, the set $\mathcal{Q}$ represents the set of syntactically valid KQL queries over a schema $s$, and a model's output $\hat{q}$ is evaluated against $\mathcal{Q}$ by Microsoft's KQL parser.

Given an NLQ, $t \in \mathcal{T}$ and optional schema context $s \in \mathcal{S}$, the main objective is for a model to produce a KQL query $\hat{q} \in \mathcal{Q}$ that is syntactically valid and as semantically faithful to the NLQ as possible. We study a translator configured by three orthogonal enhancement knobs: $\mathbf{P}$ (Prompting Scheme Enhancements),
$\mathbf{M}$ (Two-Stage Architecture with Oracle Refinement), and $\mathbf{F}$ (Fine-Tuning with LoRA). Figure~\ref{fig:knob_diagram} shows where each knob intervenes in the shared translation pipeline. Each configuration induces a mapping:
\[
F_{\mathcal{L},\mathbf{P},\mathbf{M}, \mathbf{F}} : (t,s) \rightarrow \hat{q}
\]
for a base model $\mathcal{L}$, and we evaluate $\hat{q}$ with syntax validity, schema validity, table and filter metrics, latency, and cost (Section~\ref{s:eval} defines each).
Our experiments vary these knobs to quantify the marginal contribution of each knob while keeping the base model and datasets fixed.

\subsection{Zero-Shot Prompting}

Zero-Shot prompting interacts with the model without supplying any examples or demonstrations \cite{Prompt_Engineering_Guide}. It uses fewer tokens than other prompting strategies and relies on the inference capabilities of the model itself. We begin with it to establish a baseline for how well each model generates KQL without schema context, examples, or external refinement. The prompt is given in Figure~\ref{fig:naive_zeroshot} in Appendix~\ref{sec:zeroshot-prompt}.


\subsection{NL2KQL-Inspired Prompting}
\label{sec:nl2kql-config}
\label{sssec:schema-refiner}

Zero-shot prompting establishes what an SLM can do with no support, but a model must know which tables and columns exist to avoid hallucinating identifiers. We therefore re-implement NL2KQL~\cite{nl2kql}, the state-of-the-art system for KQL generation, which grounds the model with retrieved schema context and few-shot examples through five components: a Semantic Data Catalog, Schema Refiner, Few-Shot Selector, Prompt Builder, and Query Refiner. We refer the reader to the original paper for their full descriptions.

Ours is a re-implementation with substitutions rather than an exact reproduction. Table~\ref{tab:nl2kql-substitutions} summarizes the differences. The retrieval procedure itself is unchanged. The Schema Refiner returns the top-$t$ tables and their columns by cosine similarity between the NLQ and the catalog embeddings, with $t=9$ and $v_{\text{n}}=5$ as in the original, and the Few-Shot Selector filters the FSDB by those tables before selecting examples. The Prompt Builder template and the Query Refiner's repair rules, including its 0.9 cosine-similarity threshold for identifier replacement, are also preserved. The substitutions are concentrated in four places. We build both embedding stores with Google's \texttt{text-embedding-004} rather than \texttt{text-embedding-ada-002}, at \$0.025 per million input tokens instead of \$0.10. We build the FSDB with Gemini 2.0 Flash and check syntactic and semantic correctness after generating all pairs, rather than discarding invalid ones during generation, improving efficiency and resource use. We supply $f=2$ few-shot examples rather than the original's five, matching its own 2-shot ablation. And we swap the generator across all nine evaluated models (Section~\ref{s:eval}).

\begin{table}[!t]
\centering
\footnotesize
\caption{Our re-implementation of NL2KQL~\cite{nl2kql}: components retained as-is and the four substitutions we make.}
\label{tab:nl2kql-substitutions}
\setlength{\tabcolsep}{4pt}
\renewcommand{\arraystretch}{1.15}
\begin{tabularx}{\textwidth}{@{}>{\raggedright\arraybackslash}p{2.55cm}
  >{\raggedright\arraybackslash}X
  >{\raggedright\arraybackslash}X@{}}
\toprule
\textbf{Component} & \textbf{NL2KQL} & \textbf{Ours} \\
\midrule
Semantic Data Catalog & \texttt{text-embedding-ada-002} & \texttt{text-embedding-004} \\
Schema Refiner & top-$t{=}9$ tables, $v_{\text{n}}{=}5$ values & unchanged (top 5, no values, in the two-stage system) \\
Few-Shot Selector & FSDB built by an LLM; $f{=}5$ examples; invalid pairs dropped during generation & FSDB built by Gemini 2.0 Flash; $f{=}2$ examples; correctness filtered post-generation \\
Prompt Builder & instructions, rules, few-shot examples & unchanged \\
Query Refiner & parser-driven repair, cosine threshold 0.9 & unchanged (replaced by the Oracle in the two-stage system) \\
Generator & single hosted LLM & all nine evaluated models \\
\bottomrule
\end{tabularx}
\end{table}

\PP{Error-Aware Prompting} Schema context alone does not prevent syntactic mistakes specific to KQL, so we add prompting strategies that give the model targeted tips on the errors it actually makes. The first addresses the most frequent syntactic issues reported by the KQL Parser. The second adds guidance for semantic errors such as mismatched column types and invalid table references. Because SLMs have limited inference capability, concise tips mined from the observed error distribution are more useful to them than the full KQL rule set, and they cost fewer tokens. We compare both strategies against the base NL2KQL-style prompt and adopt the one with the highest correctness.

\subsection{LoRA Fine-Tuning with LLM Distillation}




To further assess the quality of KQL queries that are generated from SLMs, we also test LoRA, a fine-tuning method that freezes pre-trained model weights from language models, and applies decompositional matrices to transformer layers within the LLM in order to alter a subset of model weights \cite{LoRA}. We choose LoRA over full fine-tuning because updating all parameters of a 6B+ model is infeasible on consumer hardware. Among parameter-efficient alternatives, LoRA is preferred over prefix tuning and adapter layers because its low-rank updates are merged into the base weights at deployment, adding no inference latency, and it matches or exceeds both in reported evaluations~\cite{LoRA}. LoRA can be mathematically defined as follows:

Given a weight matrix $\mathcal{W} \in \mathbb{R}^{d \times k}$, we decompose the update to that weight matrix into two smaller matrices such that:

\begin{center}
  $\mathcal{W'}$ = $\mathcal{W}$ + $\Delta{W}$ = $\mathcal{W}$ + $BA$
\end{center}

Where B $\in$ $\mathbb{R}^{d \times r}$, A $\in$ $\mathbb{R}^{r \times k}$, $r \ll \min(d,k)$, and both are low-rank matrices. During training time, the values of the original weight
matrix $\mathcal{W}$ remain stable while $\Delta{W}$ ($BA$) is updated. Altering low-rank matrices instead of the entire weight matrix reduces the memory requirements, 
gradient storage, and optimizer states needed to perform fine-tuning \cite{LoRA,qlora}. To train the DeepSeek Coder 6.7B Instruct, we use a synthesized dataset created through LLM Knowledge Distillation.


Using the same process as the FSDB construction above, we create a synthetic dataset of 1,000 NLQ-KQL pairs verified as syntactically and semantically correct by the KQL Parser. The teacher is Gemini 2.0 Flash, chosen for its low input and output token costs among models that still produce correct KQL. We fine-tune the student, DeepSeek Coder 6.7B Instruct, on 800 of these pairs and hold out 200 for validation.

KQL generation requires multi-step reasoning: identifying relevant tables, resolving column types, and applying operators in sequence. Distilling only final answers would transfer the teacher's outputs but not the intermediate decisions that produce them. To expose the student to reasoning signals, we augment each pair with a short chain-of-thought (CoT) explanation generated by the teacher, making the fine-tuning target a two-part sequence of explanation followed by KQL. At inference the model emits the explanation and the query, and we extract the query for scoring.

\begin{figure}[!t]
    \centering
    \begin{tcolorbox}[enhanced,
      attach boxed title to top center={yshift=-3mm,yshifttext=-1mm},
      colback=blue!5!white,
      colframe=blue!75!black,
      colbacktitle=blue!80!black,
      title=Oracle Prompting Methods,
      fonttitle=\bfseries,
      boxed title style={size=small,colframe=red!50!black},
      left=2mm, right=2mm, top=2mm, bottom=1mm, boxsep=1pt
    ]
    \footnotesize
    \textbf{Oracle for Retrieval and General Refinement:}
    \newline
    Given a Natural Language Query and a list of KQL queries, determine which of the following KQL queries is most syntactically and semantically correct:
    \newline\newline
    Natural Language Query:
    \newline
    Responses:
    \newline\newline
    Make changes as necessary to refine the best KQL query, and ensure that each column in the KQL query belongs to its respective table(s).
    Return the correct answer without explanation.
    
    \rule{\linewidth}{0.4pt}
    
    \textbf{Oracle for Retrieval and Schema Context:}
    \newline
    Given a Natural Language Query and a list of KQL queries, determine which of the following KQL queries is most syntactically and semantically correct:
    \newline\newline
    Natural Language Query:
    \newline
    Responses:
    \newline\newline
    Make changes as necessary to refine the best KQL query, and ensure that each column in the KQL query belongs to its respective table(s).
    You may use the following tables and columns:
    \newline\newline
    \{SCHEMA\}
    \newline\newline
    Return the correct answer without explanation.
    \end{tcolorbox}
    \caption{Oracle prompting templates used to guide refinement of model-generated KQL queries. The first oracle uses retrieval and general refinement, while the second incorporates schema context.}
    \label{fig:oracle-prompting}
\end{figure}


\subsection{Two-Stage Architecture}







Three simpler designs motivate a two-stage split. A single LLM end-to-end incurs prohibitive token costs at scale. A single SLM, even with strong prompting, produces too many errors to be reliable. And an SLM with a self-refinement loop is also insufficient, because SLMs lack the schema-level reasoning to judge their own outputs. We therefore assign each model the task it does best: a single DeepSeek Coder 6.7B Instruct instance, routed through LangChain, drafts the query cheaply, and an Oracle model, Gemini 2.0 Flash, applies schema-aware corrections to that draft at the final step, keeping Oracle token usage low.

\begin{figure}[!t]
  \centering
  \includegraphics[width=\columnwidth]{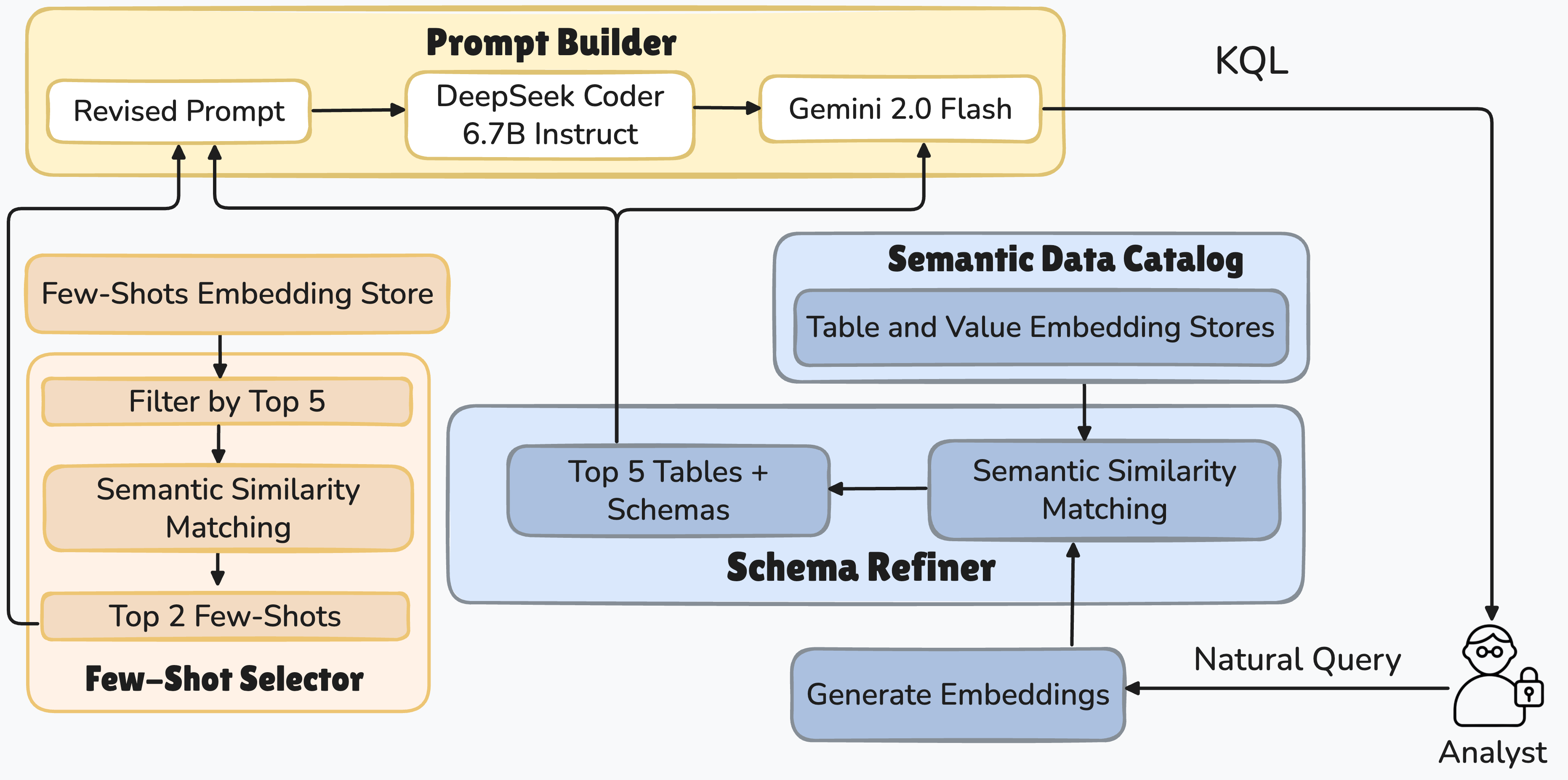}
  \caption{Two-Stage Architecture with Oracle Refinement. The NLQ is embedded to retrieve the Top 5 relevant tables, which guide the selection of Top 2 few-shot examples. These are processed by DeepSeek Coder 6.7B Instruct, and the outputs are refined by the Oracle model.}
  \label{fig:architecture_diagram}
\end{figure}

The pipeline reuses the NL2KQL components of Section~\ref{sec:nl2kql-config} with two changes. First, we reduce the Schema Refiner to the top 5 tables and omit column values, which cuts prompt tokens and narrows the SLM's opportunity to hallucinate identifiers. Table~\ref{tab:appendix-ablation-tables} reports the effect of varying this number. Second, we replace the Query Refiner with the Oracle, which repairs a broader class of errors than the original rule-based refiner and requires no hand-written repair rules. Figure~\ref{fig:architecture_diagram} shows the full pipeline and Figure~\ref{fig:pipeline-trace} traces a single NLQ through it.

The Oracle prompt is written to accept a list of candidates (Figure~\ref{fig:oracle-prompting}). In our final configuration that list holds a single SLM draft, so the Oracle acts as a refiner rather than a selector. We explore multi-candidate variants in Appendix~\ref{sec:ablation} (RQ7) and find a single candidate to be Pareto-optimal on accuracy--cost. All reported two-stage results use one SLM instance, the top 5 retrieved tables, and SLM decoding at temperature 0.2, the lowest value swept in Appendix~\ref{sec:ablation} and the one maximizing every metric there.

%

\providecommand{\tracehead}[3]{%
  \textcolor{#1!55!black}{\bfseries #2}%
  \ifx\relax#3\relax\else\ \textcolor{black!55}{#3}\fi\\[0.3mm]}
\providecommand{\tracenote}[1]{\\[0.3mm]{\normalfont\color{black!60}#1}}

\begin{figure}[!t]
\centering
\begin{tikzpicture}[
  font=\scriptsize,
  node distance=1.6mm,
  stage/.style={
    draw=#1!55!black, semithick, fill=#1!5, rounded corners=1.5pt,
    text width=0.855\linewidth, align=left, inner xsep=1.4mm, inner ysep=1.1mm
  },
  flow/.style={-{Stealth[length=1.5mm,width=1.3mm]}, semithick, draw=black!55},
]

\node[stage=bblue] (nlq) {
  \tracehead{bblue}{The analyst asks a question}{\relax}
  \ttfamily Give phishing attempts over email that did not use identity
  imitation between 20:54.33 and 21:05.12 on 2022-10-05.
};

\node[stage=ggreen, below=of nlq] (prompt) {
  \tracehead{ggreen}{Retrieval builds the prompt}{schema and examples are added to the question}
  \begin{tikzpicture}[baseline]
    \def\bh{1.9mm}
    \fill[bblue!55]  (0,0)      rectangle ++(10.9mm,\bh);
    \fill[ggreen!65] (10.9mm,0) rectangle ++(46.3mm,\bh);
    \fill[ggreen!30] (57.2mm,0) rectangle ++(16.7mm,\bh);
    \fill[rred!55]   (73.9mm,0) rectangle ++(4.0mm,\bh);
    \draw[black!45]  (0,0)      rectangle (77.9mm,\bh);
  \end{tikzpicture}\\[0.2mm]
  {\tiny
  \textcolor{bblue!55}{$\blacksquare$}\,instructions\quad
  \textcolor{ggreen!65}{$\blacksquare$}\,schema\quad
  \textcolor{ggreen!30}{$\blacksquare$}\,examples\quad
  \textcolor{rred!55}{$\blacksquare$}\,question}
  \tracenote{Retrieval keeps the 5 most relevant of 29 Defender tables; their schema
  still fills most of the prompt.}
};

\node[stage=rred, below=of prompt] (slm) {
  \tracehead{rred}{Stage 1: the SLM drafts a query}{DeepSeek Coder 6.7B Instruct}
  \ttfamily EmailEvents | where Timestamp between datetime("2022-10-05T20:54:33Z")
  .. datetime("2022-10-05T21:05:12Z")\\
  \ttfamily | where ThreatTypes contains "Phishing"\\
  \ttfamily | where DetectionMethods\textcolor{rred!75!black}{\bfseries !contains}
  "\textcolor{rred!75!black}{\bfseries Identity Imitation}"
  \tracenote{The parser rejects it: {\ttfamily !contains} is stuck to the column name
  instead of standing on its own as an operator.}
};

\node[stage=ppurple, below=of slm] (oracle) {
  \tracehead{ppurple}{Stage 2: the Oracle repairs the draft}{Gemini 2.0 Flash}
  \ttfamily | where DetectionMethods \textcolor{dkgreen}{\bfseries !contains}
  "\textcolor{dkgreen}{\bfseries IdentityImitation}"
  \tracenote{It rewrites only the broken line---separating the operator and fixing the
  value---rather than writing the query again from scratch.}
};

\node[stage=bblue, below=of oracle] (score) {
  \tracehead{bblue}{The repaired query is scored}{against the analyst's reference query}
  Syntax $=1$\quad Semantic $=1$\quad Table $=1.00$\quad
  Filter$_{\text{col}}=0.50$\quad Filter$_{\text{lit}}=0.33$
  \tracenote{It now runs and reads the right table, but it filters on
  {\ttfamily DetectionMethods} where the reference filters on {\ttfamily EmailActionPolicy}.}
};

\draw[flow] (nlq)    -- (prompt);
\draw[flow] (prompt) -- (slm);
\draw[flow] (slm)    -- (oracle);
\draw[flow] (oracle) -- (score);

\end{tikzpicture}
\caption{One analyst question traced end to end through the two-stage architecture,
with every query reproduced from an actual run. Retrieval decides prompt
size, and the retrieved schema dominates what the SLM reads. The SLM drafts a
query that is nearly right but does not parse, and the Oracle repairs only the offending
line, which keeps the second stage cheap but leaves a query that reads the correct table
yet filters on the wrong column, the gap between well-formed and faithful that the Table
and Filter scores expose.}
\label{fig:pipeline-trace}
\vspace{-2ex}
\end{figure}

The oracle operates in two modes. In \emph{general refinement}, it repairs the SLM draft using only the judge model's internal knowledge of KQL. In \emph{schema-aware refinement}, it repairs the same draft but with explicit schema context. We evaluate both modes using two prompts (Figure~\ref{fig:oracle-prompting}). The oracle is instantiated as an LLM-as-a-Judge~\cite{zheng2023judging}.

\section{Evaluation}
\label{s:eval}

We evaluate the effectiveness of our proposed design through a series of experiments conducted on a machine with an AMD EPYC 9534 64-Core Processor, an NVIDIA H100 NVL GPU, and Ubuntu 22.04.5 LTS. We address seven research questions (RQs) as follows. For all RQs, we access and query SLMs using the Huggingface Python package \cite{Huggingface}, which is not subject to input/output token rate limits. Every Gemini~2.0~Flash call is issued through the Google GenAI package \cite{PyPI}. Gemini acts as a generator in RQ1--RQ2, as the FSDB builder and distillation teacher (Section~\ref{s:methodology}), and as the Oracle in RQ5--RQ6. The OpenAI models are queried through the OpenAI API. All experiments use the same nine models: four LLMs (OpenAI GPT-5 and GPT-4o, Google Gemini 2.0 Flash, and Microsoft Phi-4) and five SLMs (Google Gemma-3-1B-IT and Gemma-3-4B-IT, Microsoft Phi-4-Mini-Instruct, Qwen2.5-7B-Instruct-1M, and DeepSeek Coder 6.7B Instruct). Our re-implemented NL2KQL pipeline (Section~\ref{sec:nl2kql-config}) is evaluated with all nine.

\begin{itemize}[leftmargin=*]
\item \textbf{RQ1}: What is the baseline KQL reasoning ability of LLMs and SLMs without enhancements?
\item \textbf{RQ2}: How well does NL2KQL perform with SLMs?
\item \textbf{RQ3}: How do different prompting schemes affect NL2KQL with SLMs?
\item \textbf{RQ4}: How effective is LoRA fine-tuning for SLM-based KQL generation?
\item \textbf{RQ5}: Can a two-stage architecture with Oracle refinement match or surpass LLM performance at lower cost?
\item \textbf{RQ6}: How well does the two-stage architecture generalize to independently authored queries over the same schema?
\item \textbf{RQ7}: What are the hyperparameter search results for different components of the two-stage architecture?
\end{itemize}

Due to space constraints, we provide results for RQ7 in Appendix~\ref{sec:ablation}.

\begin{table}[!t]
\centering
\footnotesize
\caption{Costs per 1 million tokens of input and 1 million tokens of output from LLMs and SLMs}
\label{tab:cost-analysis}
\begin{tabular}{lcc}
\toprule
Model & Input Cost & Output Cost \\
\midrule
OpenAI GPT-5 & \$1.25 & \$10.00 \\
OpenAI GPT-4o & \$2.50 & \$10.00 \\
Google Gemini 2.0 Flash & \$0.10 & \$0.40 \\
Microsoft Phi-4 & \$0.075 & \$0.30 \\
\midrule
Qwen2.5-7B-Instruct-1M & \$0.15 & \$0.15 \\
Microsoft Phi-4-Mini-Instruct & \$0.15 & \$0.15 \\
DeepSeek Coder 6.7B Instruct &  \$0.15 & \$0.15 \\
Gemma-3-4B-IT & \$0.15 & \$0.15 \\
Gemma-3-1B-IT & \$0.10 & \$0.10 \\
\bottomrule
\end{tabular}
\end{table}

\heading{Datasets} RQ1--RQ5 are evaluated using the public Defender split of Microsoft's NL2KQL Evaluation Dataset \cite{NL2KQL_Eval}, which contains 230 NLQ-KQL pairs (224 in the original paper's description) referencing all tables in the NL2KQL Defender Schema. We use the full Defender split. The dataset's Sentinel split targets a different schema and is not used anywhere in this paper. Instead of feeding NLQs directly to the models, we wrap them in different prompting strategies and compare the generated queries against the baseline pairs to measure syntactic, schema, and structural similarity. For RQ6, we use 83 NLQ-KQL pairs from the Microsoft Defender sections of Sentinel-Queries \cite{reprise99}, a public GitHub repository. We include every query whose declared data source is a Defender table present in our evaluation schema and exclude queries targeting other sources (e.g.\ Azure AD, Office 365), which keeps schema grounding comparable with RQ1--RQ5.

\heading{Evaluation Metrics}
We adapt the evaluation metrics defined in NL2KQL \cite{nl2kql}.

\PP{Syntax Score} evaluates whether an LLM-generated KQL query $\hat{q}$ is syntactically correct:
$\text{Syntax}(\hat{q}) = 1$ if $\hat{q}$ is syntactically correct, and $0$ otherwise.

\PP{Semantic Score (Schema Validity)} evaluates whether $\hat{q}$ is \emph{schema-valid} within the schema $s$:
$\text{Semantic}(\hat{q}) = 1$ if $\hat{q}$ is schema-valid, and $0$ otherwise.
Validity is checked by the semantic analyzer in Microsoft's KQL Parser~\cite{Kusto-Query-Language}, which verifies that every column referenced in $\hat{q}$ exists in the schema of the table(s) the query names. A score of 1 means the query is well-formed against the target schema and is expected to execute on Microsoft Azure Data Explorer (ADX). We keep NL2KQL's name~\cite{nl2kql} for comparability, but the metric is not semantic equivalence with the reference query. A schema-valid query can still read the wrong table or filter on the wrong column, and the Table and Filter scores below are our proxies for that intent fidelity.

The next three metrics compare the generated query $\hat{q}$ against the reference query $q$ at three levels of granularity: which tables the query reads from (Table Score), which columns it filters on (Filter$_{\text{col}}$), and which values it filters for (Filter$_{\text{lit}}$). A query can score well on the first and badly on the last, meaning it reads the right telemetry but hunts for the wrong indicator. That is exactly the failure mode we observe in \S\ref{sec:general}.

\PP{Table Score} Measures whether $\hat{q}$ reads from the right tables. It is the precision of the generated table set, gated on recall. The score is the fraction of tables in $T(\hat{q})$ that also appear in $T(q)$, but only if $\hat{q}$ covers every table the reference needs ($T(q) \subseteq T(\hat{q})$), and zero otherwise. A query that misses a required table therefore scores zero no matter what else it gets right, while the precision term penalizes a query that pulls in extra tables it does not need.

\begin{equation}
  Table(q, \hat{q})=\begin{cases}
    \frac{\vert T(q) \cap T(\hat{q}) \vert}{\vert T(\hat{q}) \vert}, & \text{if $T(q) \subseteq T(\hat{q})$}.\\
    0, & \text{otherwise}.
  \end{cases}
\end{equation}

\PP{Filter Column Score (Filter$_{\text{col}}$)} Measures whether $\hat{q}$ filters on the right columns, independent of what it compares them to. It is the Jaccard similarity between the set of columns appearing in the filter predicates of $\hat{q}$ and of $q$, where $Jaccard(a, b) = \vert a \cap b\vert / \vert a \cup b \vert$. Unlike Table Score it is symmetric, so both missing a required filter column and inventing a spurious one reduce the score. It therefore penalizes over-specification as heavily as error, and is best read as surface similarity to the reference rather than as a correctness check.

\PP{Filter Literal Score (Filter$_{\text{lit}}$)} Measures whether $\hat{q}$ filters for the right \emph{values}: the Jaccard similarity between the literal constants used in the filter predicates of $\hat{q}$ and of $q$. These are the IP addresses, file hashes, action types, and time bounds that determine what the query actually matches.

To make the three concrete, consider the reference query in Figure~\ref{fig:kql-example-two} and the generated query in Figure~\ref{fig:kql-example-three}. Both read only \texttt{EmailEvents}, so $T(q)=T(\hat{q})=\{\texttt{EmailEvents}\}$ and Table Score is $1$. The reference filters on \{\texttt{Timestamp}, \texttt{ThreatTypes}, \texttt{EmailActionPolicy}\}, but the generated query filters on only \{\texttt{Timestamp}, \texttt{ThreatTypes}\}, giving Filter$_{\text{col}} = 2/3 = 0.67$, because it dropped the anti-phishing policy condition. The literals diverge further. The generated query searches for both \texttt{"Phish"} and \texttt{"Phishing"} and omits the policy string, so Filter$_{\text{lit}}$ falls below Filter$_{\text{col}}$. This is the typical ordering across our results and the reason we report all three, since Table Score alone would rate this query perfect.

We add two efficiency measures:

\PP{Avg. Latency} Average end-to-end model inference time per NLQ (s/query), measured from prompt assembly to final string, including network/API overhead. For the two-stage architecture, latency is the sum of the SLM drafting time and the Oracle's refinement time per NLQ.

\PP{Cost} Total cost (USD) to process 230 NLQs from the NL2KQL Defender Evaluation Dataset, accounting for all input and output tokens per model. For hosted LLMs we use published vendor API prices. Open-weight SLMs have no list price, since the weights are a free download, but serving them is not, so we adopt Cloudflare Workers AI pricing \cite{cloudflare2024workersai} as a representative rate for hosting an SLM for a SOC (\$0.10 per million tokens up to 3B parameters, \$0.15 from 3.1B to 8B). Section~\ref{s:discussion} explains why this upper-bounds on-premise serving and makes our SLM figures conservative. Table~\ref{tab:cost-analysis} presents the cost per 1M tokens in USD.

\heading{Hyperparameters} To train SLMs with LoRA-based Parameter-Efficient Fine-Tuning (PEFT), we keep the following hyperparameters stable for DeepSeek Coder 6.7B Instruct in RQ4: one training epoch, batch size of 5 per device, learning rate of 0.0002, weight decay of 0.001, maximum gradient norm of 0.3, and warmup ratio of 0.03. The optimizer used is \texttt{paged\_adamw\_32bit}. For supervised fine-tuning, we vary the LoRA parameters and select the combination minimizing validation loss. Specifically, we test alpha ($\alpha$) values of 2, 4, and 8; rank ($r$) values of 1, 2, and 4; and dropout values of 0.1 and 0.2.

\heading{Embedding Model} The primary embedding model used in the following RQs is Google's \texttt{text-embedding-004}, as described in Section~\ref{sssec:schema-refiner}.

\subsection{RQ1: Baseline LLM and SLM Performance}



We first establish the baseline KQL reasoning capability of the nine models without any enhancements. No contextual information is supplied through the prompt, so the models must produce a KQL query from the NLQ alone using the Zero-Shot prompt of Figure~\ref{fig:naive_zeroshot}. Results are shown in Table~\ref{tab:nl2kql-llm-slm-eval}.

Without schema enhancements, LLMs generate syntactically correct KQL queries but struggle with semantic accuracy. OpenAI's GPT-5, GPT-4o and Google Gemini 2.0 Flash outperform Microsoft Phi-4 on every accuracy metric, identifying relevant tables more effectively. Accuracy and latency are decoupled, however. GPT-4o and Gemini~2.0~Flash answer in roughly a second, whereas GPT-5 averages 35.7~s per NLQ, or roughly 35$\times$ slower, despite leading every intent-fidelity metric (schema validity, table score, and Filter$_{\text{col}}$). This already signals that the strongest model is not automatically the right one to deploy in an interactive SOC workflow. SLMs, in contrast, perform poorly on their own. Across all five tested, table score and Filter$_{\text{col}}$ remain below 0.1. While SLMs can produce syntactically valid KQL, they are largely unable to generate semantically correct queries. The failure is driven by hallucinations of table and column names, or by references to columns that do not belong to the queried table's schema.

\begin{table*}[!t]
\centering
\small
\caption{Evaluation of LLM and SLM prompting configurations (including NL2KQL) using syntax, semantic, table score, filtering accuracy, and efficiency metrics. Latency is the average time per NLQ (s/query). Cost is the total USD to run all 230 queries, based on token prices per million.}
\label{tab:nl2kql-llm-slm-eval}
\resizebox{\textwidth}{!}{%
\begin{tabular}{llccccccc}
\toprule
\multicolumn{2}{l}{\textbf{Model and Configuration}}  & \textbf{Syntax}  &  \textbf{Semantic} & \textbf{Table} &\textbf{Filter$_{\text{col}}$} &\textbf{Filter$_{\text{lit}}$} & \textbf{Latency} & \textbf{Cost} \\
\midrule
\multirow{2}{*}{GPT-5} & Zero-Shot & 0.900 & 0.700 & 0.700 & 0.404 & 0.520 & 35.711 & \$0.267 \\
& NL2KQL & 0.930 & 0.861 & 0.283 & 0.114 & 0.377 & 53.670 & \$2.018 \\
\midrule
\multirow{2}{*}{GPT-4o} & Zero-Shot & 0.970 & 0.274 & 0.452 & 0.289 & 0.542 & 1.033 & \$0.136 \\
& NL2KQL & 0.961 & 0.878 & 0.696 & 0.385 & 0.549 & 10.008 & \$2.998 \\
\midrule
\multirow{2}{*}{Gemini 2.0 Flash} & Zero-Shot & 0.973 & 0.282 & 0.383 & 0.266 & 0.540 & 1.095 & \$0.009 \\
& NL2KQL & 0.883 & 0.812 & 0.716 & 0.487 & 0.527 & 2.385 & \$0.107 \\
\midrule
\multirow{2}{*}{Microsoft Phi-4} & Zero-Shot & 0.845 & 0.029 & 0.126 & 0.061 & 0.407 & 2.126 & \$0.012 \\
& NL2KQL & 0.694 & 0.516 & 0.604 & 0.342 & 0.424 & 11.733 & \$0.127 \\
\midrule
\multirow{2}{*}{Microsoft Phi-4-Mini-Instruct} & Zero-Shot & 0.623 & 0.007 & 0.044 & 0.026 & 0.389 & 1.232 & \$0.004 \\
& NL2KQL & 0.664 & 0.279 & 0.510 & 0.332 & 0.420 & 5.398 & \$0.127 \\
\midrule
\multirow{2}{*}{Gemma-3-1B-IT} & Zero-Shot & 0.443 & 0.007 & 0.000 & 0.000 & 0.294 & 1.373 & \$0.005 \\
& NL2KQL & 0.771 & 0.332 & 0.321 & 0.220 & 0.253 & 4.286 & \$0.084 \\
\midrule
\multirow{2}{*}{Gemma-3-4B-IT} & Zero-Shot & 0.741 & 0.000 & 0.004 & 0.000 & 0.373 & 1.793 & \$0.008 \\
& NL2KQL & 0.803 & 0.446 & 0.629 & 0.317 & 0.495 & 6.859 & \$0.127 \\
\midrule
\multirow{2}{*}{Qwen2.5-7B-Instruct-1M} & Zero-Shot & 0.826 & 0.003 & 0.026 & 0.009 & 0.479 & 0.947 & \$0.005 \\
& NL2KQL & 0.641 & 0.413 & 0.419 & 0.310 & 0.401 & 6.258 & \$0.126 \\
\midrule
\multirow{2}{*}{DeepSeek Coder 6.7B Instruct} & Zero-Shot & 0.810 & 0.027 & 0.065 & 0.022 & 0.450 & 3.328 & \$0.009 \\
& NL2KQL & 0.862 & 0.710 & 0.626 & 0.398 & 0.496 & 9.455 & \$0.127 \\
\bottomrule
\end{tabular}%
}
\end{table*}

\subsection{RQ2: SLMs with NL2KQL}


We next measure how LLMs and SLMs perform within the NL2KQL system, as described in Section~\ref{sec:nl2kql-config}. We reimplement NL2KQL's architecture with some substitutions and test it on the four LLMs and five SLMs. These results are shown in Table~\ref{tab:nl2kql-llm-slm-eval}.

Every model improves over its zero-shot configuration on schema validity, and that is where the gain is concentrated. It rises from near 0 to between 0.27 and 0.71 across the five SLMs, because the retrieved schema context supplies the table and column names the models otherwise hallucinate. The other metrics do not move uniformly. Syntax falls for four of nine models, most steeply for Qwen2.5-7B (0.826 to 0.641) and Phi-4 (0.845 to 0.694), and GPT-5 loses table score (0.700 to 0.283) and Filter$_{\text{col}}$ (0.404 to 0.114) even as its schema validity rises to 0.861.

The two GPT models lead on syntax and schema validity while trailing Gemini~2.0~Flash on table score and Filter$_{\text{col}}$. This split follows in part from what those metrics measure rather than from query quality alone. Syntax and schema validity judge a query on its own terms, whereas Table Score's precision term and both filter scores are set comparisons against a terse, human-authored reference. A correct but unrequested predicate lowers the score exactly as an incorrect one does, so a model that elaborates beyond the reference is penalized for doing so. Provenance may compound this. Our FSDB is generated by Gemini~2.0 Flash (Section~\ref{sec:nl2kql-config}), so in this configuration Gemini alone receives few-shot examples drawn from its own output distribution and reproduces their conventions most faithfully. That is a self-consistency advantage on similarity metrics, and it says nothing about capability. This is why we match systems on schema validity, which needs no reference, when we compare cost in RQ5.

The improvement is also not free. Cost and average latency rise for every model, driven by the larger prompt at inference time. We leave the latency and cost entries blank for the NL2KQL row reported by Tang et al.\ (Table~\ref{tab:nl2kql-eval}), as these values were not reported in \cite{nl2kql}.

\begin{figure}[!t]
\centering
\begin{minipage}{0.95\linewidth}
\begin{lstlisting}[language=SQL,
    basicstyle=\ttfamily\scriptsize,
    keywordstyle=\color{blue}\bfseries,
    commentstyle=\color{gray}\itshape,
    stringstyle=\color{teal},
    showstringspaces=false,
    frame=single,
    caption = {The KQL query above gives a list of phishing attempts over email that did not use identity imitation between 20:54.33 and 21:05.12 on 2022-10-05.},
    captionpos=b,
    label={fig:kql-example-two}]
EmailEvents
| where Timestamp between(datetime(" 2022-10-05T20:54:33Z") .. datetime("2022-10-05T21:05:12Z"))
| where ThreatTypes has "Phish"
| where EmailActionPolicy != "Anti-phishing user impersonation"
\end{lstlisting}
\end{minipage}
\vspace{-3ex}
\end{figure}

\begin{figure}[!t]
\centering
\begin{minipage}{0.95\linewidth}
\begin{lstlisting}[language=SQL,
    basicstyle=\ttfamily\scriptsize,
    keywordstyle=\color{blue}\bfseries,
    commentstyle=\color{gray}\itshape,
    stringstyle=\color{teal},
    showstringspaces=false,
    frame=single,
    caption={This KQL query was generated by DeepSeek Coder 6.7B Instruct with the NL2KQL configuration for the same NLQ as in Figure~\ref{fig:kql-example-two}. Although similar to the ground truth, the generated query is syntactically and semantically incorrect due to incorrect usage of the KQL keyword: has\_any.},
    captionpos=b,
    label={fig:kql-example-three}]
EmailEvents
| where Timestamp between (datetime(2022-10-05 20:54:33) .. datetime(2022-10-05 21:05:12))
| where has_any(ThreatTypes, "Phish", "Phishing")
| project Timestamp, NetworkMessageId, SenderMailFromAddress, RecipientEmailAddress, Subject, ThreatTypes, DetectionMethods
\end{lstlisting}
\end{minipage}
\vspace{-3ex}
\end{figure}

Figures~\ref{fig:kql-example-two} and~\ref{fig:kql-example-three} illustrate a representative failure case. The ground-truth query in Figure~\ref{fig:kql-example-two} correctly applies \texttt{has} in infix form, while the SLM-generated query in Figure~\ref{fig:kql-example-three} misuses \texttt{has\_any} as a prefix function call, a recurring syntactic error that also violates KQL semantics. This class of error feeds the two largest categories in Table~\ref{tab:syntax_tbl}, stray markdown (``Unexpected: `'', 28.6\%) and incomplete fragments (24.6\%), which together motivate the error-aware prompting of RQ3.

At the same token cost, DeepSeek Coder 6.7B Instruct leads every other SLM on syntax, schema validity, and both filter scores, and matches the best of them on table score. We therefore build on it under the NL2KQL configuration in the remaining RQs.

\subsection{RQ3: Different Prompt Schemes}

We now examine how different prompting schemes affect NL2KQL with SLMs. To revise the prompt, we first need to know which errors DeepSeek Coder 6.7B Instruct actually makes under the NL2KQL configuration, so we run Microsoft's KQL code parser \cite{Kusto-Query-Language} over its outputs. Table~\ref{tab:syntax_tbl} in Appendix~\ref{sec:table-refs} lists the most common syntactic errors.

From this table, the most frequent syntax errors were ``Unexpected:`'' and ``The incomplete fragment is unexpected.'', followed by missing delimiters such as ``Expected:~;'' and ``Expected:~)\,''. These issues commonly arise when the model mixes markdown fragments, emits partial SQL-like text, or produces malformed parentheses in operators such as \texttt{between}. By enforcing a single KQL output, constraining timestamp expressions, and requiring operators to appear only in infix form, Alternative Prompt~\#1 reduces these incomplete or ill-formed fragments. For this reason, we used the Alternative Prompt~\#1 strategy as outlined in Appendix~\ref{sec:prompt-specs}. Table~\ref{tab:nl2kql-eval} outlines the updated results from these tests.



Alternative Prompt~\#1 raises the syntax, semantic, and Filter$_{\text{col}}$ scores (Table~\ref{tab:nl2kql-eval}). Re-running the parser confirms the intended mechanism. Total syntax errors fall from 427 to 210 (Table~\ref{tab:syntax_tbl_additional}), and the two classes the prompt targets, stray markdown (``Unexpected: `'') and incomplete fragments, account for most of that drop.

We then apply the same parser-driven procedure to the semantic score. As Table~\ref{tab:semantic-tbl} shows, generic complaints remain the largest single classes under Alternative Prompt~\#1 (``Expected:~;'' 6.2\%, incomplete fragments 5.3\%). The errors it newly exposes are improperly referenced columns, namely ``Column name expected.'' (5.4\%) and undefined identifiers such as \texttt{FileName} (3.4\%). Both are absent from the base configuration's top five, and both arise from hallucination or from columns outside the queried table's schema. Alternative Prompt~\#2 (Appendix~\ref{sec:prompt-specs}) adds rules aimed at these cases, and its results are reported in Table~\ref{tab:nl2kql-eval}.


  \begin{table*}[!t]
    \centering
    \footnotesize
    \caption{Evaluation of NL2KQL and various model configurations on the Defender Dataset.}
    \label{tab:nl2kql-eval}
    \resizebox{\textwidth}{!}{%
    \begin{tabular}{ll ccccccc}
      \toprule
    \textbf{Category} & \textbf{Model Configuration} & \textbf{Syntax} & \textbf{Semantic} & \textbf{Table} & \textbf{Filter$_{\text{col}}$} & \textbf{Filter$_{\text{lit}}$} & \textbf{Latency} & \textbf{Cost}\\
    \midrule
    Baseline & NL2KQL (reported by Tang et al.)\textsuperscript{\dag} & 0.993 & 0.972 & 0.834 & 0.694 & 0.605 & -- & -- \\
    \midrule
    \multirow{3}{*}{\makecell[l]{Prompting\\Strategies}}
    & NL2KQL + DeepSeek (Original) & 0.862 & 0.710 & 0.626 & 0.398 & 0.496 & 9.455 & \$0.127 \\
    & NL2KQL + DeepSeek (Revised \#1) & 0.924 & 0.753 & 0.618 & 0.467 & 0.456 & 13.550 & \$0.240 \\
    & NL2KQL + DeepSeek (Revised \#2) & 0.869 & 0.600 & 0.550 & 0.424 & 0.442 & 14.346 & \$0.257 \\
    \midrule
    \multirow{4}{*}{\makecell[l]{Advanced\\Configurations}}
    & DeepSeek (Supervised LoRA Fine-Tuning) & 0.911 & 0.726 & 0.564 & 0.452 & 0.475 & 13.900 & \$0.260 \\
    & DeepSeek (CoT LoRA Fine-Tuning) & 0.909 & 0.716 & 0.577 & 0.470 & 0.477 & 13.910 & \$0.260 \\
    & Two-Stage (General Refinement) & 0.830 & 0.537 & 0.594 & 0.463 & 0.497 & 18.267 & \$0.266 \\
    & Two-Stage (Schema Context) & 0.987 & 0.906 & 0.659 & 0.537 & 0.562 & 12.315 & \$0.213 \\
    \bottomrule
    \end{tabular}%
    }
    \vspace{0.5ex}
    \raggedright\scriptsize\textsuperscript{\dag}Reported scores from the original NL2KQL paper~\cite{nl2kql}, not our reproduction. The original reports one evaluation over its combined Defender and Sentinel benchmark, so this row is not strictly comparable with our Defender-only numbers. Our reimplementation with Gemini~2.0 Flash achieves 0.883/0.812 (syntax/semantic; see Table~\ref{tab:nl2kql-llm-slm-eval}). We attribute the gap primarily to our use of a different embedding model (\texttt{text-embedding-004} vs.\ \texttt{text-embedding-ada-002}), a different few-shot generation model, and two few-shot examples rather than five.
    \end{table*}


  

Alternative Prompt~\#2 improves on Alternative Prompt~\#1 on no metric, and DeepSeek Coder 6.7B Instruct performs considerably worse under it. Only syntax (0.869 vs.\ 0.862) and Filter$_{\text{col}}$ (0.424 vs.\ 0.398) end up marginally above the original NL2KQL prompt, and schema validity falls below it. The parser output explains why. As the third column of Table~\ref{tab:semantic-tbl} shows, undefined columns and variables persist, and become more frequent, even though the prompt explicitly instructs the model to respect the supplied schemas. Stating schema rules in the prompt is therefore not the same as being able to check them, and SLMs cannot independently verify whether their own queries are schema-valid.

\subsection{RQ4: LoRA Fine Tuning}



We next ask whether LoRA fine-tuning improves KQL generation for DeepSeek Coder 6.7B Instruct. We first perform a hyperparameter search that minimizes the cross-entropy validation set loss. As shown in Table~\ref{tab:alpharank}, the configuration $\alpha=8$, $r=1$, and LoRA dropout = 0.2 produced a validation set loss of 0.0105, and we insert the model fine-tuned under it into the NL2KQL system. The results are shown in Table~\ref{tab:nl2kql-eval} under Advanced Configurations.



Fine-tuning helps only against the weakest baseline. Relative to the original NL2KQL prompt it raises schema validity from 0.710 to 0.726 (+0.016) and syntax from 0.862 to 0.911. But Alternative Prompt~\#1, the best RQ3 strategy and the one the fine-tuned model is deployed with, reaches 0.753, above both fine-tuned variants. Fine-tuning also adds nothing at inference time. The fine-tuned configurations run at \$0.260 against \$0.240 for Alternative Prompt~\#1, the strategy they are built on.

Supervised fine-tuning on answers alone may not be sufficient for reasoning SLMs. In phishing email detection, small models fine-tuned on emails and labels alone collapse relative to the same models fine-tuned on explanation-augmented targets, with Llama-3.2-3B-Instruct falling from 0.928 to 0.219 F1 \cite{Improving_Phishing_Email_Detection}.

We therefore also attempt a CoT approach, augmenting each NLQ--KQL pair with a short explanation from the teacher (Gemini~2.0 Flash) of how the KQL result was derived. This gives the model the reasoning behind a query in addition to the query itself. We select LoRA parameters by the same criterion as above (Table~\ref{tab:alpharank_two}) and report the results in Table~\ref{tab:nl2kql-eval} under Advanced Configurations. CoT fine-tuning is comparable to supervised fine-tuning in producing syntactically and semantically correct KQL queries, but it does not surpass any of the prompting strategies.

LoRA fine-tuning on 800 synthetic pairs is insufficient for schema-aware KQL generation, even with CoT distillation, and the negative result is informative. Fine-tuning transfers surface patterns such as operator usage and query structure, but not the schema-grounding judgment needed to select valid columns from a large, evolving schema, which RQ3 already showed SLMs cannot supply on their own. If that judgment cannot be cheaply baked into SLM weights, it must be supplied at inference time, which is what the two-stage architecture does in RQ5.

\subsection{RQ5: Two-Stage Architecture}




\PP{Setup} We evaluate the two-stage architecture of Section~\ref{s:methodology}. A DeepSeek Coder 6.7B Instruct draft is produced with the best prompting strategy from RQ3 and decoded at temperature 0.2, then refined by an Oracle LLM (Google Gemini 2.0 Flash, chosen for its low token costs among LLMs that still yield correct KQL). One retrieval setting changes with the architecture. The drafting stage uses the reduced Schema Refiner of Section~\ref{s:methodology} (top 5 tables, no enumerated column values), whereas the NL2KQL rows of RQ2--RQ3 retain the original top-9-with-values retrieval. The two-stage prompt is therefore shorter per query than the RQ3 prompt it inherits, which is why its \emph{total} cost and latency across both stages can fall below the single-stage RQ3 row. Appendix~\ref{sec:ablation} sweeps this choice. To isolate the Oracle's contribution we compare its two prompts (Figure~\ref{fig:oracle-prompting}). Under \emph{general refinement} it revises the draft using only its own knowledge of KQL, and under \emph{schema-aware refinement} it additionally receives the schema context retrieved by the Schema Refiner. Results are reported in Table~\ref{tab:nl2kql-eval}.

\PP{Results} The two-stage architecture with schema context achieves the highest syntax (0.987) and semantic (0.906) scores of any configuration we report, exceeding both the revised prompting strategies of RQ3 and every model in the NL2KQL reimplementation of RQ2. The choice of Oracle prompt matters substantially. Without schema context, general refinement reaches only 0.830 syntax and 0.537 semantic, below the revised prompting strategy it is applied to (0.924 / 0.753). The Oracle is therefore not simply cleaning up SLM output. It requires the retrieved schema to do so, and without it can degrade a draft it has no means to verify. On cost, the schema-aware configuration runs 230 queries for \$0.213. That is roughly 2$\times$ our pure-Gemini NL2KQL reimplementation (\$0.107), but 9.5$\times$ cheaper than our GPT-5-based NL2KQL reimplementation (\$2.018) and 14$\times$ cheaper than our GPT-4o-based one (\$2.998). We note that our two-stage system does not match the NL2KQL results originally reported by Tang et al.\ in terms of syntax and semantic score.

\PP{Interpretation} Cost ratios depend entirely on which baseline is chosen, so we fix an accuracy bar first and compare cost only among systems that clear it. Of the configurations in Tables~\ref{tab:nl2kql-llm-slm-eval} and~\ref{tab:nl2kql-eval}, only NL2KQL+GPT-4o and NL2KQL+GPT-5 land within 0.05 schema-valid of our system, and they are also the two most expensive systems we run. Every cheaper baseline falls at least 0.09 below it. The 9.5$\times$ and 14$\times$ figures are therefore accuracy-matched rather than chosen for the largest ratio. The counterpart holds too. An operator willing to give up 0.094 schema validity can run our NL2KQL+Gemini~2.0~Flash reimplementation at half our cost, and we do not claim to dominate it on price.

We claim cost at matched accuracy rather than speed. At 12.315~s per NLQ we sit between the two accuracy-matched baselines (GPT-4o at 10.008~s, GPT-5 at 53.670~s) and well behind that Gemini reimplementation at 2.385~s. Two sequential model calls are inherent to the design, so latency is predictable rather than low. Every query costs one draft plus one refinement, with no confidence-triggered retries. The comparison is also not like-for-like in either direction, since our SLM runs locally and unbatched while the hosted baselines run on latency-optimized infrastructure. Where response time is the binding constraint rather than schema validity, the Gemini reimplementation is the better choice.

That +0.094 gain over the pure-LLM reimplementation does not come from adding schema context to a cheap LLM, since Gemini~2.0 Flash already receives schema context in the NL2KQL pipeline. It comes from decoupling drafting from refinement. The SLM produces a draft under a code-focused prompt, and the Oracle applies schema-aware edits to that draft rather than regenerating from scratch. This division of labor appears to offer a different inductive bias than a single Gemini pass that must both draft and self-correct in one shot.

\subsection{RQ6: Generalization to Independently Authored Queries}
\label{sec:general}

Having established the two-stage architecture on the NL2KQL Defender evaluation dataset, we now ask whether its performance holds on queries written independently of that benchmark. We evaluate on the 83 Sentinel-Queries pairs described above, applying the same prompting strategy as the two-stage architecture with schema context. This holds the schema fixed, since these are the same Defender tables as RQ1--RQ5. What changes is the query distribution, because practitioners wrote these queries for their own use and they differ from the benchmark pairs in phrasing and operator mix.

As shown in Table~\ref{tab:nl2kql-final-sentinel-eval}, the two-stage architecture achieves high syntax (0.964) and semantic (0.831) scores on the independently authored Sentinel-Queries dataset. However, literal filtering essentially fails ($\text{Filter}_{\text{lit}}=0.100$), and column filtering is weak (0.267). These high-level scores are encouraging but mask a critical deployment risk. In real SOC threat hunting, literal values such as IP addresses, file hashes, and timestamps are often the most important filter in a query. A query that selects the right table and columns but filters on the wrong IP address is operationally useless and potentially dangerous, as it silently returns no results for the actual threat.

The drop from the in-domain semantic score (0.906) to 0.831 cannot be blamed on an unfamiliar schema, since these queries read the same tables. It reflects phrasing instead. Retrieval matches query text against catalog embeddings, so practitioner phrasing retrieves a less apt set of five tables and the Oracle validates against columns never supplied. The literal-level drop is sharper still, as the values a practitioner filters for are rarely recoverable from the schema at all. The architecture thus transfers its structural KQL competence to independently authored queries, but column and literal precision remain sensitive to the query distribution its retrieval was tuned on.


\begin{table}[!t]
\centering
\footnotesize
\caption{Evaluation of the Two-Stage Architecture on the Sentinel-Queries Dataset. Lat.\ = average latency (s/query); Cost = total USD to run all 83 queries.}
\label{tab:nl2kql-final-sentinel-eval}
\setlength{\tabcolsep}{2pt}
\begin{tabular}{ccccccc}
\toprule
\textbf{Syntax} & \textbf{Semantic} & \textbf{Table} & \textbf{Filter$_{\text{col}}$} & \textbf{Filter$_{\text{lit}}$} & \textbf{Lat.} & \textbf{Cost} \\
\midrule
0.964 & 0.831 & 0.429 & 0.267 & 0.100 & 12.37 & \$0.078 \\
\bottomrule
\end{tabular}
\end{table}

\section{Discussion \& Limitations}
\label{s:discussion}

\PP{What the SLM Cost Figures Represent} Pricing open-weight SLMs at a serverless rate (Table~\ref{tab:cost-analysis}) may seem at odds with our emphasis on self-hosting, but the two are compatible. The weights we evaluate are a few gigabytes and free to download, so counting the download alone makes every SLM cost zero and the comparison against hosted APIs vacuous. The real expense is serving them. An organization running an SLM for its SOC pays for accelerator capacity, power, and the overhead of keeping that endpoint available, and those costs depend on its GPU inventory, request volume, and batching efficiency. Rather than assume one hardware amortization, we use a published per-token rate for the same service, which places all nine models in the same unit.

\PP{Direction of the Bias} This choice is conservative. A commercial rate includes the provider's margin and its idle capacity, so it upper-bounds what an organization pays on hardware it already owns, where the marginal cost of a query approaches the electricity to run it. Hosted LLM prices allow no such reduction, since the API price \emph{is} the cost. A cheaper SLM rate, including zero for a fully self-hosted draft model, would only widen the gap we report, so the 9.5$\times$ and 14$\times$ reductions over NL2KQL+GPT-5 and NL2KQL+GPT-4o are lower bounds. The exception is the all-Gemini baseline, which already costs less than our system (Section~\ref{s:eval}). A lower SLM rate narrows that gap rather than reversing it.

\PP{One Model in Three Roles} Gemini~2.0 Flash generates the FSDB examples, teaches the distilled student, and serves as the Oracle, inviting a self-preference objection~\cite{zheng2023judging}. Two things bound it. The Oracle refines a single DeepSeek draft rather than selecting among candidates, which is the setting in which such bias is measured, and no reported metric is judge-supplied. Validity comes from the KQL Parser and the filter scores from human-authored references. A shared blind spot remains possible, and separating the three roles across model families is the clean test we leave to future work.

\PP{Practical Implications for SOC Deployment} Taken together, our results point to a deployment pattern rather than a single model choice. That pattern is local or private SLMs for generation, a narrow low-cost LLM for schema-aware judgment and refinement, and strict parser validation before any query is executed. It keeps drafting cheap and latency predictable, at exactly one draft and one refinement per query, while confining the expensive model to the one step where its schema reasoning is actually needed.

\PP{Generalizability to Other Query Languages} The framework extends beyond KQL to EQL~\cite{Elastic_Query_Language_Documentation} and SPL~\cite{Search_Manual} by re-instantiating three components rather than redesigning the pipeline:

\begin{itemize}[leftmargin=*,itemsep=0.2em,topsep=0.3em]
\item \textbf{The schema source.} We treat EQL indexes and SPL index--sourcetype pairs as ``tables'' and build field lists from Elasticsearch mappings or Splunk knowledge objects, then retrieve top-$t$ sources and top-$v$ fields as in Section~\ref{sssec:schema-refiner}. Few-shot selection is unchanged, and only the serialization of examples into native syntax differs.
\item \textbf{The validator.} Microsoft's KQL Parser is replaced by an EQL dry-run parse or the SPL parser, each augmented with per-stage field-existence checks.
\item \textbf{The metrics.} Table scores become index or index--sourcetype scores, and the two filter scores remain Jaccard over fields and literals, still penalizing out-of-schema fields.
\end{itemize}

We have not evaluated these ports empirically, and the prompt-level guardrails that suppress KQL parser errors (Appendix~\ref{sec:prompt-specs}) would need to be re-mined from each target language's own error distribution. Transfer to a schema the Semantic Data Catalog has never embedded is likewise untested. Both our evaluations run against the Defender schema, and RQ6 varies who wrote the queries, not what they query.

\section{Conclusion}
\label{s:conclusion}
We present the first systematic study of SLMs for NLQ-to-KQL translation, covering prompting enhancements, LoRA fine-tuning with rationale distillation, and a two-stage SLM-Oracle architecture. The two-stage design attains the highest syntax and schema-valid scores of any configuration we report, at 9.5$\times$ and 14$\times$ lower cost than the NL2KQL+GPT-5 and NL2KQL+GPT-4o reimplementations, the only baselines within 0.05 schema-valid of it.

Two findings qualify this result. Architecture mattered more than fine-tuning: LoRA on 800 synthetic pairs, with or without rationale distillation, did not surpass targeted prompting, indicating that schema grounding is better supplied at inference time than compiled into SLM weights. Generalization is also structural rather than complete. On independently authored queries over the same Defender schema the system preserves syntax and schema validity but loses literal-level precision, selecting the right telemetry while filtering for the wrong values. Whether this holds for a genuinely unseen schema remains open. Closing that gap, likely through retrieval over observed column values rather than schemas alone, is the clearest path to SLM-generated queries that are directly actionable in a SOC.

\begin{credits}
\subsubsection{\ackname}
We thank the anonymous reviewers for their valuable feedback. This material is based upon work supported by the National Science Foundation (NSF) under Award Nos. 2339483, 2530655, and 2622986, and by the Commonwealth Cyber Initiative (CCI).

\subsubsection{\discintname}
The authors have no competing interests to declare that are relevant to the content of this article.
\end{credits}

\bibliographystyle{splncs04}
\bibliography{bib-files/hassan,bib-files/hassan-misc}

\appendix
\section{Zero-Shot Prompt}
\label{sec:zeroshot-prompt}

\begin{figure}[!t]
    \centering
    \begin{tcolorbox}[enhanced,
        attach boxed title to top center={yshift=-3mm,yshifttext=-1mm},
        colback=blue!5!white,
        colframe=blue!75!black,
        colbacktitle=blue!80!black,
        title=Zero-Shot Prompting,
        fonttitle=\bfseries,
        boxed title style={size=small,colframe=red!50!black}, ,
      left=2mm, right=2mm, top=2mm, bottom=1mm, boxsep=1pt
    ]
    \footnotesize
    You are a programmer using the Kusto Query Language with Microsoft Defender.
    Generate a KQL query that answers the following request:
    \newline\newline
    \{NLQ\}
    \newline\newline
    Return only the KQL code without any explanation.
    \end{tcolorbox}
    \caption{Zero-Shot prompt used to evaluate how SLMs generate KQL queries}
    \label{fig:naive_zeroshot}
\end{figure}

The Zero-Shot Prompt in Figure~\ref{fig:naive_zeroshot} supplies minimal context to assess the SLM's baseline KQL reasoning without schema or example support; it corresponds to RQ1 (Section~\ref{s:eval}).

\section{Alternative Prompts}
\label{sec:prompt-specs}
We introduce two revised prompt templates targeting common errors we found using Microsoft's KQL Parser; see Figure~\ref{fig:alternative-prompting}. Alternative Prompt 1 enforces strict schema discipline and prevents the model from generating undefined Defender-style fields. It specifies the correct infix usage of some operators such as ``between'', ``has\_any'', and ``contains''. Alternative Prompt 2 builds on this by requiring the model to choose a valid starting table and to reference only those columns defined in that table (or in explicitly joined tables). Both templates state their rules compactly, but they are applied on top of the same retrieved schema and few-shot blocks as the base NL2KQL prompt, so the added guardrails lengthen rather than shorten the prompt---which is what drives the cost increase reported for the revised strategies in Table~\ref{tab:nl2kql-eval}. This complements the NL2KQL setup described in Section~\ref{sec:nl2kql-config}.

\begin{figure}[!t]
    \centering
    \begin{tcolorbox}[enhanced,attach boxed title to top center={yshift=-3mm,yshifttext=-1mm},
        colback=blue!5!white,colframe=blue!75!black,colbacktitle=blue!80!black,
        title=Alternative Prompting Methods,fonttitle=\bfseries,
        boxed title style={size=small,colframe=red!50!black}]
        \scriptsize
        \textbf{Alternative Prompt \#1:}
    
        You are a cybersecurity expert generating syntactically valid KQL for Microsoft Defender/Sentinel/USX. Return only a single KQL query without any explanation.
        \par\smallskip
        Use the following structures to describe a timestamp: between(datetime(...) .. datetime(...)) or Timestamp < ago(7d)
        \par\smallskip
        Use \texttt{contains}, \texttt{has}, \texttt{has\_any}, \texttt{has\_all}, \texttt{in}, \texttt{in\~}, and \texttt{between} operators ONLY in infix form. 
        For example: \texttt{Column has\_any ("a","b")}, \texttt{Timestamp between (datetime(...) .. datetime(...))}.        \par\smallskip
        Never invent new tables, columns, or values. Use only tables/columns/data types in:
        \par\smallskip
        \{SCHEMA\}
        \par\smallskip
        Use only values stated in the NLQ.
        \par\smallskip
        Examples:
        \par\smallskip
        \{EXAMPLES\}
        \par\smallskip
        \{NLQ\}
        \newline
        \rule{\linewidth}{0.4pt}
        \textbf{Alternative Prompt \#2:}
    
        You are a cybersecurity expert generating syntactically valid KQL for Microsoft Defender/Sentinel/USX. Return only a single KQL query inside a \texttt{kusto} code block without any explanation.
        \par\smallskip
        Use ONLY the following structures to describe timestamps: \texttt{Column between (datetime(...) .. datetime(...))} or \texttt{Timestamp < ago(7d)}.
        \par\smallskip
        Use \texttt{contains}, \texttt{has}, \texttt{has\_any}, \texttt{has\_all}, \texttt{in}, \texttt{in\~}, and \texttt{between} operators ONLY in infix form. 
        For example: \texttt{Column has\_any ("a","b")}, \texttt{Timestamp between (datetime(...) .. datetime(...))}.
        \par\smallskip
        Never invent new tables, columns, or values. Use only tables/columns/data types explicitly listed in:
        \par\smallskip
        \{SCHEMA\}
        \par\smallskip
        If the NLQ mentions a field not present in the schema, ignore that part of the request.
        \par\smallskip
        Negation must use \texttt{not(...)}. Do NOT use the "!" operator anywhere.
        \par\smallskip
        The query must start with a table name (no leading "|") and follow a single pipeline structure.
        \par\smallskip
        Use only values stated in the NLQ.
        \par\smallskip
        Examples:
        \par\smallskip
        \{EXAMPLES\}
        \par\smallskip
        \{NLQ\}
        \newline        
        \label{box:rq3-box2}
    \end{tcolorbox}
    \caption{Alternative prompting templates used to reduce common KQL errors.}
    \label{fig:alternative-prompting}
\end{figure}

\section{Table References}
\label{sec:table-refs}
Table~\ref{tab:syntax_tbl} lists the most common syntax errors that were associated with the DeepSeek NL2KQL Configuration (RQ2). In total, there were 427 syntax errors from the SLM-generated KQL queries across five iterations. The most prevalent error type was ``Unexpected: `'', followed by ``The incomplete fragment is unexpected''. While the third most common error, ``Expected: ;'', is not particularly useful in fixing the KQL queries due to its generic nature, the fourth error, ``Expected: )'' (7.5\%), suggests issues with improper parenthesis matching in function calls or logical expressions.

Table~\ref{tab:semantic-tbl} tracks the semantic errors across all three prompting configurations of RQ3. In the base NL2KQL configuration (860 errors), the dominant classes are stray markdown and incomplete fragments, alongside type errors such as ``A value of type timespan expected'', which indicates difficulty constructing temporal expressions; these patterns inform Alternative Prompt~\#1. After that prompt, the total falls to 793 and the residual errors shift toward improperly referenced columns: ``Column name expected'' and undefined identifiers such as \texttt{FileName} (3.4\%), which are consistent with SLM hallucination or with assuming a column belongs to a table it does not. Alternative Prompt~\#2 does not help; the total rises to 1185 and ``Column name expected'' grows from 5.4\% to 8.6\%, while the \texttt{FileName} diagnostic holds roughly constant in absolute terms (3.4\% of 793 versus 2.3\% of 1185, or about 27 occurrences either way). This is the appendix-level view of the regression reported in RQ3.

\begin{table}[!t]
    \centering
    \scriptsize
    \caption{Most common syntax errors: DeepSeek NL2KQL Configuration over five iterations (Total: 427).}
    \label{tab:syntax_tbl}
    \begin{tabular}{lc}
        \toprule
        \textbf{Syntax Error} & \textbf{Percentage of Errors} \\
        \midrule
        Unexpected: ` & 28.6\% \\
        The incomplete fragment is unexpected. & 24.6\% \\
        Expected: ; & 13.3\% \\
        Expected: ) & 7.5\%\\
        Expected: ( & 4.9\% \\
        \bottomrule
    \end{tabular}
\end{table}

\begin{table}[!t]
    \centering
    \scriptsize
    \caption{Most common diagnostics from the KQL Parser's semantic analyzer across the three RQ3 configurations, as a percentage of that configuration's total. The analyzer reports whatever blocks validation, so its output mixes residual syntax faults with true schema faults; the shift between the two is the effect the alternative prompts target. A dash means the diagnostic was not in that configuration's top five.}
    \label{tab:semantic-tbl}
    \renewcommand{\arraystretch}{1.15}
    \setlength{\tabcolsep}{8pt}
    \begin{tabularx}{\textwidth}{@{}>{\raggedright\arraybackslash}Xccc@{}}
        \toprule
        & \textbf{NL2KQL} & \textbf{Alt.\ \#1} & \textbf{Alt.\ \#2} \\
        \textbf{Parser Diagnostic} & (860) & (793) & (1185) \\
        \midrule
        Unexpected: ` & 14.2\% & -- & -- \\
        The incomplete fragment is unexpected. & 12.2\% & 5.3\% & 5.1\% \\
        Expected: ; & 6.6\% & 6.2\% & 7.7\% \\
        Expected: ) & 3.7\% & -- & -- \\
        A value of type timespan expected. & 2.7\% & -- & -- \\
        Column name expected. & -- & 5.4\% & 8.6\% \\
        A value of type string or dynamic expected. & -- & 3.5\% & -- \\
        The name 'FileName' does not refer to any known column, table, variable or function. & -- & 3.4\% & 2.3\% \\
        Expected: , & -- & -- & 4.8\% \\
        \bottomrule
    \end{tabularx}
\end{table}

 Table~\ref{tab:syntax_tbl_additional} shows the most common syntax errors that were found after the changes in prompting in the DeepSeek NL2KQL Configuration. The number of syntax errors significantly decreased with the new prompting strategy.
\begin{table}[!t]
    \centering
    \scriptsize
    \caption{Most common syntax errors after revised prompting strategy (Total: 210).}
    \label{tab:syntax_tbl_additional}
    \begin{tabular}{lc}
        \toprule
        \textbf{Syntax Error} & \textbf{Percentage of Errors} \\
        \midrule
        Expected: ; & 23.3\% \\
        The incomplete fragment is unexpected. & 20.0\% \\
        Missing: " & 11.0\% \\
        Unexpected: \textbackslash & 10.5\% \\
        Missing expression & 9.0\% \\
        \bottomrule
    \end{tabular}
\end{table}

\section{RQ7: Hyperparameter Search Studies}
\label{sec:ablation}



In the following section, we vary different components of the two-stage architecture and assess how this affects the overall results of the system. We note that these experiments constitute a \emph{hyperparameter search} rather than a traditional ablation study: we do not remove architectural components to measure their individual contribution; instead, we sweep parameter values to identify settings that best balance cost and correctness. The four main areas of testing include the \textbf{number of SLMs}
that are used within the two-stage architecture system, the different \textbf{temperature values} used within the two-stage architecture system, the \textbf{number of tables} fed into the SLM and Oracle LLM prompt, and the \textbf{LoRA alpha and rank values} used when fine-tuning the SLM.

The number-of-SLMs and temperature sweeps were run at different points while the pipeline was still being tuned, with the prompting and retrieval settings finalized afterwards. Their absolute scores therefore sit within 0.015 of the final system reported in Table~\ref{tab:nl2kql-eval} and within that margin of each other where the two sweeps overlap (both contain the single-SLM, temperature-0.2 setting). We report them for the trends they establish, which the later refinements do not disturb; because they predate the final configuration, their absolute values are not comparable with Table~\ref{tab:nl2kql-eval} and we do not count them among the configurations reported there. The number-of-tables sweep was re-run on the final configuration, which is why its Top 5 row (Table~\ref{tab:appendix-ablation-tables}) reproduces Table~\ref{tab:nl2kql-eval} exactly.

\begin{table}[!t]
    \centering
    \footnotesize
    \caption{Evaluation of Two-Stage Architecture with varying numbers of top tables supplied. The bolded Top 5 row is the configuration used for the two-stage system throughout the paper; its scores are identical to those reported in Table~\ref{tab:nl2kql-eval}.}
    \label{tab:appendix-ablation-tables}
    \resizebox{\columnwidth}{!}{%
    \begin{tabular}{ll ccccccc}
      \toprule
    \textbf{Category} & \textbf{Model Configuration} & \textbf{Syntax} & \textbf{Semantic} & \textbf{Table} & \textbf{Filter$_{\text{col}}$} & \textbf{Filter$_{\text{lit}}$} & \textbf{Latency} & \textbf{Cost}\\
    \midrule
    \multirow{5}{*}{\makecell[l]{Two-Stage Arch.\\(Top $t$ tables)}}
    & Top 1 & 0.983 & 0.849 & 0.584 & 0.515 & 0.544 & 10.175 & \$0.150 \\
    & Top 3 & 0.983 & 0.887 & 0.637 & 0.523 & 0.550 & 11.361 & \$0.183 \\
    & \textbf{Top 5} & \textbf{0.987} & \textbf{0.906} & \textbf{0.659} & \textbf{0.537} & \textbf{0.562} & \textbf{12.315} & \textbf{\$0.213} \\
    & Top 7 & 0.989 & 0.886 & 0.680 & 0.545 & 0.552 & 13.342 & \$0.244 \\
    & Top 9 & 0.978 & 0.877 & 0.703 & 0.550 & 0.561 & 14.256 & \$0.274 \\
    \bottomrule
    \end{tabular}%
    }
    \end{table}

\begin{figure}[H]
  \centering
  \begin{minipage}[t]{0.49\columnwidth}
    \centering
    \includegraphics[width=\linewidth]{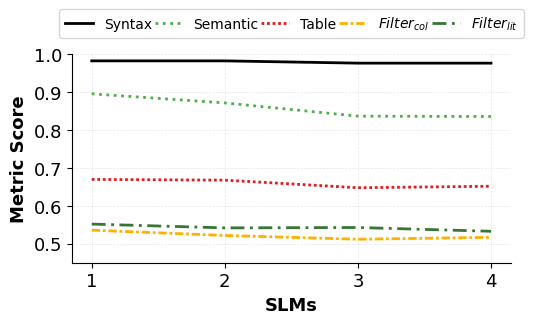}
    \caption{Number of SLMs vs. Metric Scores}
    \label{fig:ablation_slms}
  \end{minipage}\hfill
  \begin{minipage}[t]{0.49\columnwidth}
    \centering
    \includegraphics[width=\linewidth]{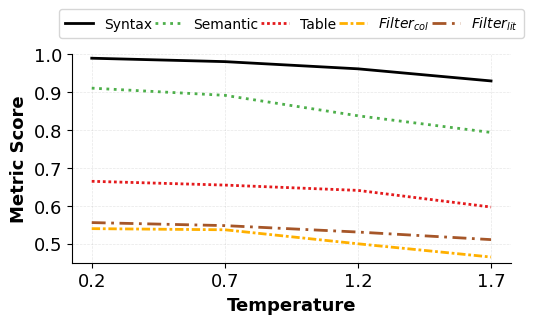}
    \caption{Temperature vs. Metric Scores}
    \label{fig:ablation_temperature}
  \end{minipage}
  \vspace{-2ex}
\end{figure}

\PP{Number of SLMs}
The number of SLMs that are utilized in the two-stage architecture solution can potentially introduce a tradeoff between metric scores and latency. A greater number of SLMs utilized in the system often means that the system will likely incur higher latency.
However, the greater the number of SLMs that are utilized in the two-stage architecture solution, the more plausible responses are generated. In this hyperparameter search study, we determine what effect the number of SLMs utilized has on responses while keeping other
components (i.e., temperature) consistent. Figure~\ref{fig:ablation_slms} outlines the metric scores as the number of SLMs utilized in the system increases. Increasing the number of SLMs utilized in the system reduces all metric scores in the entire two-stage system. When the number of SLMs utilized in the two-stage system is one, the syntax score is 0.983, the semantic score is
0.896, the table score is 0.67, the $Filter_{col}$ score is 0.536, and the $Filter_{lit}$ score is 0.552. However, when the number of SLMs utilized in the two-stage system increases to two, 
the syntax score stays the same, but the semantic score is reduced to 0.872, the table score is reduced to 0.66, the $Filter_{col}$ score is reduced to 0.522, and the $Filter_{lit}$ score is reduced to 0.542. This reduction trend continues as the number of SLMs utilized
in the system continue to increase as shown by the figure. Therefore, the best number of SLMs to utilize in the two-stage system is one.

This sweep fixes the generator count for the rest of the paper: every configuration we report uses exactly one SLM instance, and the remaining effort goes into improving that single generator rather than adding more. The $N=1$ scores above (0.983 syntax, 0.896 semantic) come from the preliminary build noted at the start of this section; with the prompting and Top-5 schema-refiner settings finalized, the same single-SLM design reaches 0.987 and 0.906 in Table~\ref{tab:nl2kql-eval}. That gap is the gain from those later refinements, not from a change in the number of generators.

This result is counterintuitive: one would expect best-of-$N$ selection to improve with more candidates, as the oracle has a richer pool to choose from. The reason it does not is a quality--diversity tradeoff in oracle/judge model behavior. Each additional SLM instance generates candidates at the same temperature, but the marginal candidates are not independent draws of comparable quality---they tend to be lower-quality variations that introduce more hallucinated columns and malformed operators. The oracle must then discriminate among a larger set of candidates with a lower average quality floor, which degrades its selection accuracy. In effect, adding candidates dilutes the pool rather than enriching it. This suggests a practical design principle for multi-stage generation systems: when the judge model has bounded discrimination capacity, a single high-quality generator paired with strong prompting outperforms a diverse but noisy candidate set.

\begin{table}[!t]
    \centering
    \scriptsize
    \setlength{\tabcolsep}{3.5pt}
    \begin{minipage}[t]{0.49\textwidth}
        \centering
        \caption{Hyperparameter search validation losses (Supervised Fine-Tuning).}
        \label{tab:alpharank}
        \begin{tabular}{cccc}
            \toprule
            \textbf{Alpha} & \textbf{Rank} & \makecell{\textbf{LoRA}\\\textbf{Dropout}} & \makecell{\textbf{Validation}\\\textbf{Loss}} \\
            \midrule
            8 & 1 & 0.2 & 0.010477 \\
            8 & 1 & 0.1 & 0.010488 \\
            8 & 2 & 0.2 & 0.010500 \\
            8 & 2 & 0.1 & 0.010526 \\
            8 & 4 & 0.2 & 0.010882 \\
            8 & 4 & 0.1 & 0.010882 \\
            4 & 2 & 0.1 & 0.017084 \\
            4 & 2 & 0.2 & 0.018105 \\
            4 & 1 & 0.2 & 0.018485 \\
            4 & 1 & 0.1 & 0.018823 \\
            4 & 4 & 0.1 & 0.019539 \\
            4 & 4 & 0.2 & 0.019622 \\
            2 & 2 & 0.1 & 0.026694 \\
            2 & 2 & 0.2 & 0.026819 \\
            2 & 1 & 0.1 & 0.029698 \\
            2 & 4 & 0.1 & 0.029962 \\
            2 & 4 & 0.2 & 0.030118 \\
            2 & 1 & 0.2 & 0.030627 \\
            \bottomrule
        \end{tabular}
    \end{minipage}\hfill
    \begin{minipage}[t]{0.49\textwidth}
        \centering
        \caption{Hyperparameter search validation losses (CoT Fine-Tuning).}
        \label{tab:alpharank_two}
        \begin{tabular}{cccc}
            \toprule
            \textbf{Alpha} & \textbf{Rank} & \makecell{\textbf{LoRA}\\\textbf{Dropout}} & \makecell{\textbf{Validation}\\\textbf{Loss}} \\
            \midrule
            8 & 1 & 0.2 & 0.010473 \\
            8 & 1 & 0.1 & 0.010489 \\
            8 & 2 & 0.2 & 0.010507 \\
            8 & 2 & 0.1 & 0.010515 \\
            8 & 4 & 0.2 & 0.010867 \\
            8 & 4 & 0.1 & 0.010871 \\
            4 & 2 & 0.1 & 0.017208 \\
            4 & 2 & 0.2 & 0.018153 \\
            4 & 1 & 0.2 & 0.018505 \\
            4 & 1 & 0.1 & 0.018788 \\
            4 & 4 & 0.1 & 0.019508 \\
            4 & 4 & 0.2 & 0.019609 \\
            2 & 2 & 0.1 & 0.026689 \\
            2 & 2 & 0.2 & 0.026858 \\
            2 & 1 & 0.1 & 0.029772 \\
            2 & 4 & 0.1 & 0.029944 \\
            2 & 4 & 0.2 & 0.030118 \\
            2 & 1 & 0.2 & 0.030347 \\
            \bottomrule
        \end{tabular}
    \end{minipage}
\end{table}

\PP{Different Temperature Values}
We tested the two-stage architecture system under four different temperature values: 0.2, 0.7, 1.2, 1.7. These temperature values are meant to represent low, moderate, and high temperature values respectively. When configured within SLMs, low temperatures provide more deterministic
code while higher temperatures introduce more randomness and creativity in token generation. In this hyperparameter search study, we study how varying the temperature of the SLMs can affect the quality of KQL queries that are produced from the proposed system. Figure~\ref{fig:ablation_temperature} outlines the metric scores as the temperature utilized in the SLM within the two-stage system varies. When the temperature
is set to 0.2, the syntax score is 0.99, the semantic score is 0.911, the table score is 0.665, the $Filter_{col}$ score is 0.54, and the $Filter_{lit}$ score is 0.556. When the temperature is set to 0.7, the syntax score is reduced to 0.981, the semantic score is reduced to 0.892, the table score is reduced to 0.655, the $Filter_{col}$ score is reduced to 0.537, and the $Filter_{lit}$ score is reduced to 0.548.
When the temperature is set to 1.2, the syntax score is 0.962, the semantic score is 0.838, the table score is 0.641, the $Filter_{col}$ score is 0.5, and the $Filter_{lit}$ score is 0.531. Lastly, when the temperature is set to 1.7, the syntax score is 0.93, the semantic score is 0.794, the table score is 0.597, the $Filter_{col}$ score is 0.465, and the $Filter_{lit}$ score is 0.511. There does appear to be a clear
relationship between temperature and performance, with all metric scores decreasing as the temperature increases.

\PP{Different Number of Tables} We tested the two-stage architecture when it receives varying numbers of top tables ranging from 1 to 9 tables. The goal in providing fewer tables is to reduce total costs and latency in utilizing the system. However, we also seek to determine how varying the number of tables given to the SLM and Oracle LLM affects other metrics. These results are shown in Table~\ref{tab:appendix-ablation-tables}.

The syntax scores remain relatively stable across varying numbers of tables supplied, ranging from 0.978 to 0.989. The semantic score generally increases as the number of tables supplied increases, reaching a peak of 0.906 with five tables before slightly declining with seven and nine tables. The table score shows a clear upward trend, increasing from 0.584 with one table to 0.703 with nine tables. However, this improvement comes at a cost: both latency and average cost increase substantially, with latency growing from 10.175 seconds (1 table) to 14.256 seconds (9 tables), and cost increasing from \$0.15 to \$0.274. To reach a good balance between cost, efficiency, and metrics, we choose our ideal number of tables to supply to the system as five, which achieves the highest semantic score (0.906) while maintaining reasonable latency (12.315 seconds) and cost (\$0.213).

\PP{Different Alpha and Rank Values}
In Table~\ref{tab:alpharank} and Table~\ref{tab:alpharank_two}, we report the validation losses obtained from each combination of parameters while training DeepSeek Coder 6.7B Instruct on NLQ-KQL pairs: Table~\ref{tab:alpharank} for the Supervised Fine-Tuning case and Table~\ref{tab:alpharank_two} for the CoT case. The two regimes rank the configurations near-identically and differ by less than $10^{-4}$ at every setting, which is itself informative: adding rationale tokens to the training target barely moves the loss surface, consistent with the RQ4 finding that CoT distillation does not improve generated KQL.


\end{document}